\documentclass[fleqn,usenatbib]{mnras}

\usepackage{newtxtext,newtxmath}

\usepackage[T1]{fontenc}

\DeclareRobustCommand{\VAN}[3]{#2}
\let\VANthebibliography\thebibliography
\def\thebibliography{\DeclareRobustCommand{\VAN}[3]{##3}\VANthebibliography}

\usepackage{graphicx}	
\usepackage{amsmath}	
\usepackage{amsfonts}

\title[L-BASS]{L-BASS: A project to produce an absolutely calibrated 1.4 GHz sky map. ~~ II - Technical Description of the System.  }

\author[D. P. Zerafa  et al.]{
D. P. Zerafa,$^{1}$
P. N. Wilkinson,$^{1}$
C. J. Radcliffe,$^{2}$
J. P. Leahy,$^{1}$
I. W. A. Browne$^{1}$
P. J. Black,$^{1}$ \thanks{E-mail: phillip.black@postgrad.manchester.ac.uk}
\\
$^{1}$Jodrell Bank Centre for Astrophysics, 
Department of Physics and Astronomy, The University of Manchester, Oxford Road, Manchester, M13 9PL, UK\\
$^{2}$Phase2 Microwave Ltd, Unit 1a, Boulton Rd, Pin Green Ind. Est., Stevenage, SG1 4QX, UK\\
}

\date{Accepted 21 May 2025.}

\pubyear{2025}

\usepackage{amssymb}	

\begin{document}
\label{firstpage}
\pagerange{\pageref{firstpage}--\pageref{lastpage}}
\maketitle

\begin{abstract}
L-BASS is an instrument designed to make radiometric temperature measurements of the sky with an absolute accuracy of better than $0.1$ K at 1.4 GHz. This will be achieved in two steps: first by measuring the sky temperature relative to that of the North Celestial Pole, using two horn-based antennas, and second with the sky antenna replaced with a calibrated cryogenic load to measure the absolute brightness temperature of the North Celesial Pole.  Here we describe the design of the L-BASS two-antenna system and report on laboratory measurements to establish its performance at component and sub-system level. 

\end{abstract}




\section{Introduction}

As described in \citet[][hereafter Paper I]{Paper1} the primary scientific motivations for L-BASS (the L-Band All Sky Survey) are: i) to establish the zero-level of all-sky maps at 1.4~GHz made with dish-type radio telescopes \citep[e.g.][]{Reich2004,Calabretta2014,Wolleben2021}; ii) to provide independent evidence to test the reality of a new isotropic steep spectrum radio background proposed by the ARCADE-2 team \citep{Fixsen2011,Seiffert2011}

Paper I also describes and illustrates the dual horn-based telescope we have constructed to make the necessary absolute total power measurements, accurate to $0.1$~K. All observations will be made within the radio astronomy protected band 1400~MHz to 1427~MHz and given this small ($\approx 2\%$) bandwidth the horns can be smooth-walled with a multi-stepped throat section generating the modes required to tailor the beam shape. The horn design is described in Sec.~\ref{sec:horn}. Below the throat a septum polarizer, described in Sec.~\ref{sec:polarizer}, provides both hands of circular polarization. One output of the polarizer is fed into the receiver while the other is used to inject a single-frequency, known as continuous wave (CW), calibration signal. The performance of the composite antennas consisting of the horns plus the polarizers is described in Sec.~\ref{sec:polandhorntests}. To allow the horns to tip over a wide elevation range, flexible RF cables, 4.04-m long, connect the polarizers to the inputs of the receiver.  These cables are a critical part of the system and much effort has gone into characterizing their behaviour as described in Sec.~\ref{sec:4-m cables}.

The receiving system is described Sec.~\ref{sec:receiver}. The principal module is a continuous comparison direct conversion radiometer whose architecture is similar to that of {\it WMAP} \citep{Bennett2013} and {\it Planck} LFI \citep{2010A&A...520A...4B}; all the RF amplification is at the observation frequency.  The receiver is housed in a temperature-controlled box mounted on the telescope structure and its outputs are connected by a pair of $\sim$55~m RF cables to a digital spectrometer situated in a nearby building.
The spectrometer independently samples both outputs from the receiver and covers the protected band.  In software the sampled bandwidth is restricted further to 1400 - 1425 MHz, giving us a bandpass with 456 frequency channels each $\approx 55$~kHz wide.

\section{horn design and performance}
\label{sec:horn}

The primary requirement for the horns is that the beam sidelobes should be as low as possible, certainly better than $-40$\,dB with respect to the main axis, in order to suppress unwanted pickup from the local environment and the Sun. As shown in Paper I,  we also surrounded L-BASS with a large ground screen to further suppress the pickup of stray radiation.  

Given the second of our scientific goals, the FWHM beam-width of the horns would, ideally, be similar to the 12\degr\ of ARCADE-2; however practicalities of scale and cost constrained the design to have a significantly larger beam. It is well known that the best performing horn design in terms of bandwidth and sidelobes is the corrugated horn, but these are expensive to manufacture for 21~cm wavelength. Simpler and cheaper alternatives are possible for operation over a narrow, $\sim 2 $$\%$, bandwidth. For ease of manufacture we have settled on a modified Potter type horn. The basic Potter horn has a circular cross section with a single step in diameter at the throat but \cite{Leech2011} showed that the beam could be tailored by utilising several tapered steps. We adopted this technique for the L-BASS horn design. 

A suite of software tools is required to simulate the performance of different design choices. Our principal tool was the Mician Microwave Wizard mode matching program \citep{Microwave_Wizard}  which has a comprehensive set of routines for antenna and horn design. Early design work revealed that realising a FWHM beamwidth approaching that of ARCADE-2 would require a horn almost 5~m long which was unrealistic in terms of mechanical mounting and manufacturing cost. A smaller 2.5~m length horn, with an aperture diameter of 0.78~m, was eventually chosen; this realised a FWHM beamwidth of $\approx 23^\circ$.

\begin{figure*} 
    \centering
    \includegraphics[width=10.0cm]{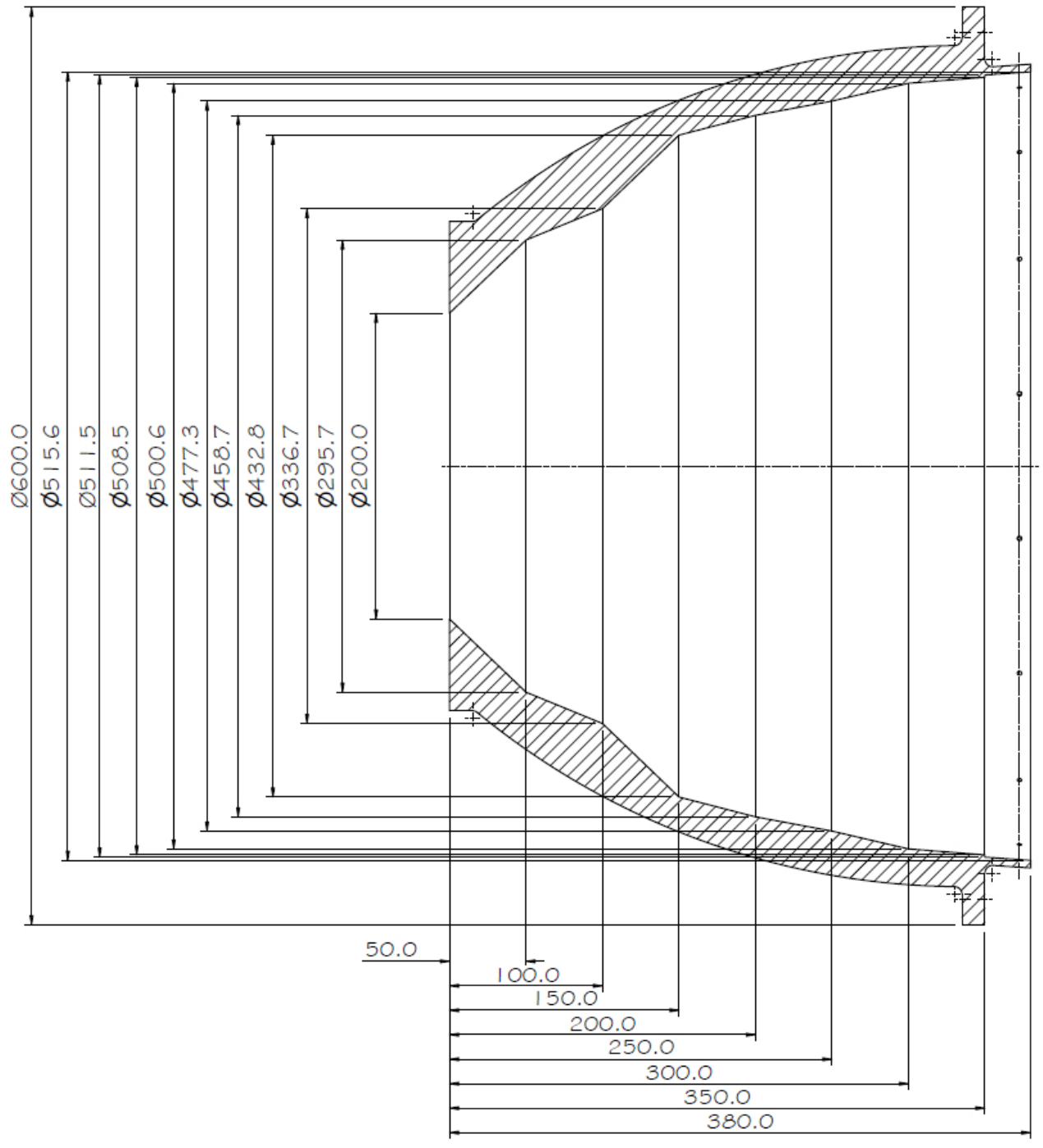}
    \caption{A cross section of the tapered stepped throat part of the L-BASS horn.}
    \label{stepped-throat}
\end{figure*}

Initially a throat design with 11 steps in diameter was investigated. In order to reduce complexity the on-axis distance between the steps was constrained to be 50~mm. A 2~m length of constant angle horn (cone) was then added to the model to give the required beamwidth. All the step-radii were the optimised with the constraint that the radii had to increase from the throat to the aperture. Although the system was intended to be used with circular polarization, for simplicity linear E fields were used for optimisation. The  beamwidth at $\phi=0\degr$ and $\phi=90\degr$  was constrained to be equal, and the sidelobes were constrained to be less than --44~dB when the off-axis angle, $\theta$, was between 37\degr\ and 50\degr\, less than --48~dB between 51\degr\ and 72\degr\  and less than --60~dB between 73\degr\ and 180\degr. The error function was calculated at seven frequencies across the 27~MHz wide protected band. While this optimisation was successful it was apparent that some of the steps in opening angle were not contributing to the overall performance because there was minimal change in radius between adjacent steps. A 7-step design was found to give virtually the same performance as one with more steps and this was the solution adopted. A cross-section through the final design for the stepped part of the horn is shown in Figure~\ref{stepped-throat}.

\begin{figure*}   
\centering
\includegraphics[scale=0.7]{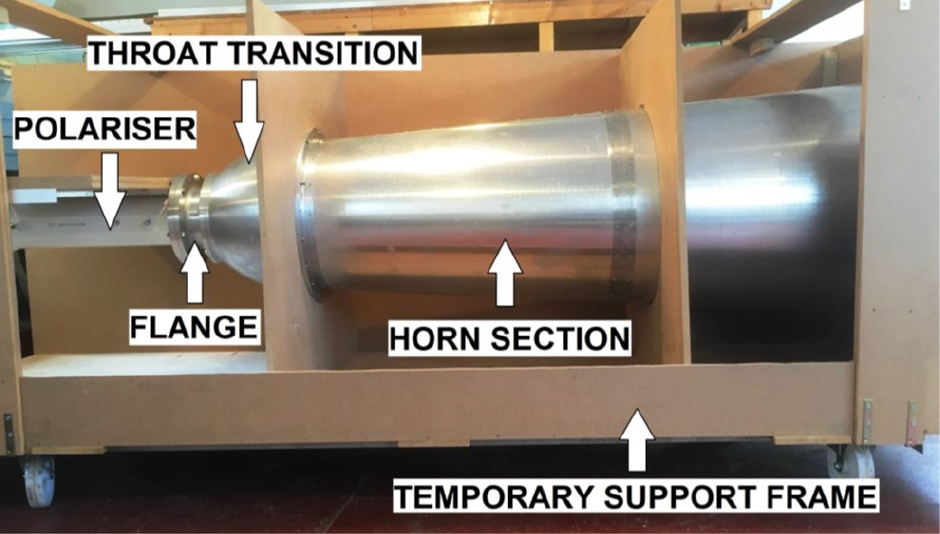}
\caption{An antenna mounted in a temporary support frame prior to testing in the laboratory; its overall length is 3.36m.} 
\label{antenna_in_frame} 
\end{figure*}

The subtle internal shapes required in the throat sections were separately machined out of solid aluminium blocks by an external manufacturer. The constant opening angle cone sections of the horns were manufactured in the Jodrell Bank Observatory (JBO) workshop by rolling sheet aluminium of thickness 2mm. The horns and throat sections were mated together at JBO and the polarizers attached to the throat via a matching flange (see Figure~\ref{antenna_in_frame}). The whole horn/polarizer assemblies were then mounted in welded aluminium cradles for lab testing and eventual installation on the telescope structure.

\begin{figure*} 
    \centering
    \includegraphics[width=10.0cm]{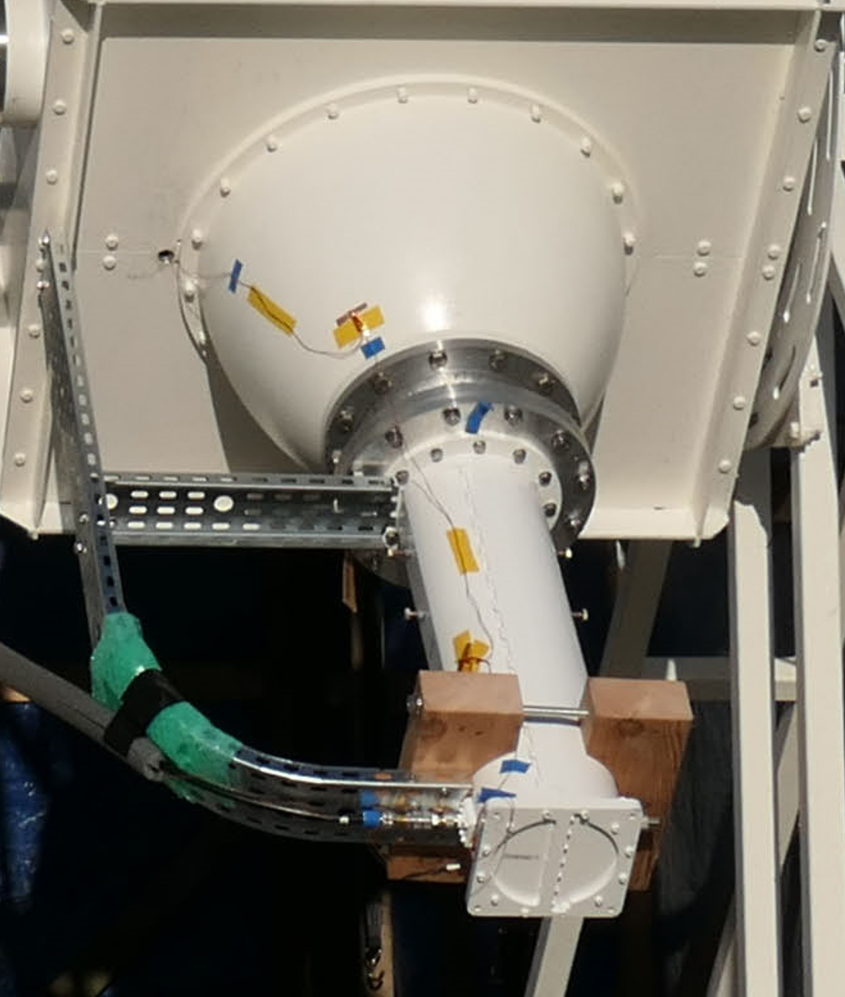}
    \caption{The back end of one of the antennas mounted on the telescope structure showing the outside of the stepped throat section, the flange and the septum polarizer. The initially constrained section of the 4.04-m coaxial cable carrying the signal to one input of the receiver can be seen, together with some of the temperature sensors of the 1-Wire network.}
    \label{horn-and-polariser}
\end{figure*}

\begin{figure*}
    \centering
    \includegraphics[scale=0.45]{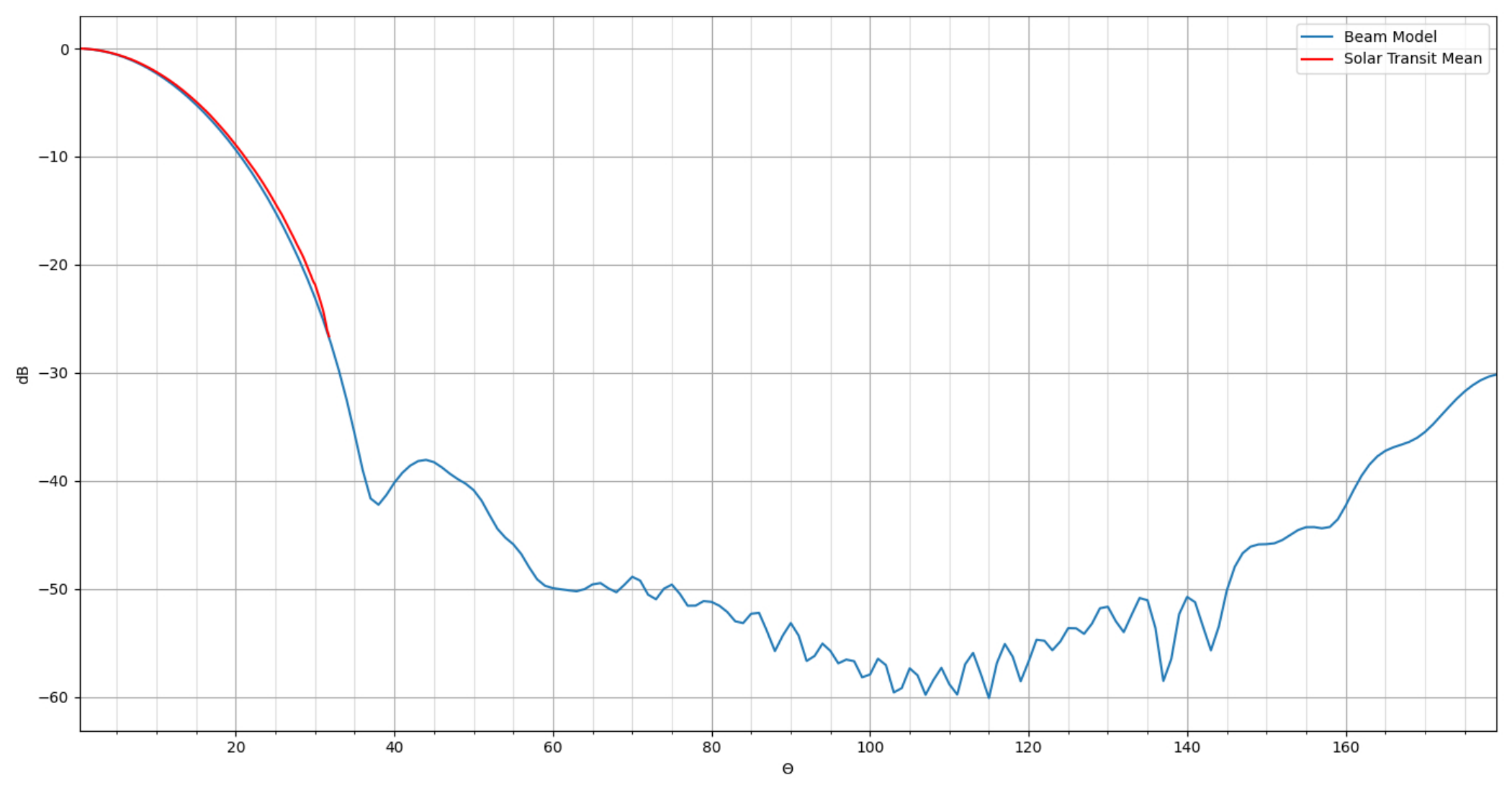}
    \caption{The simulated co-polar beam (blue) and the beam as measured by solar transit (red). The solar transit data plotted are the average of 8 separate transit observations taken between March 2022 and October 2024 and enable the beam to be traced down to $\sim$ -25 dB of the peak.  The simulated beam is symmetric both in azimuth $\theta$ and about the boresight direction. }
    \label{beam_vs_solar}
\end{figure*}

\begin{figure*}
    \centering
    \includegraphics[scale=0.45]{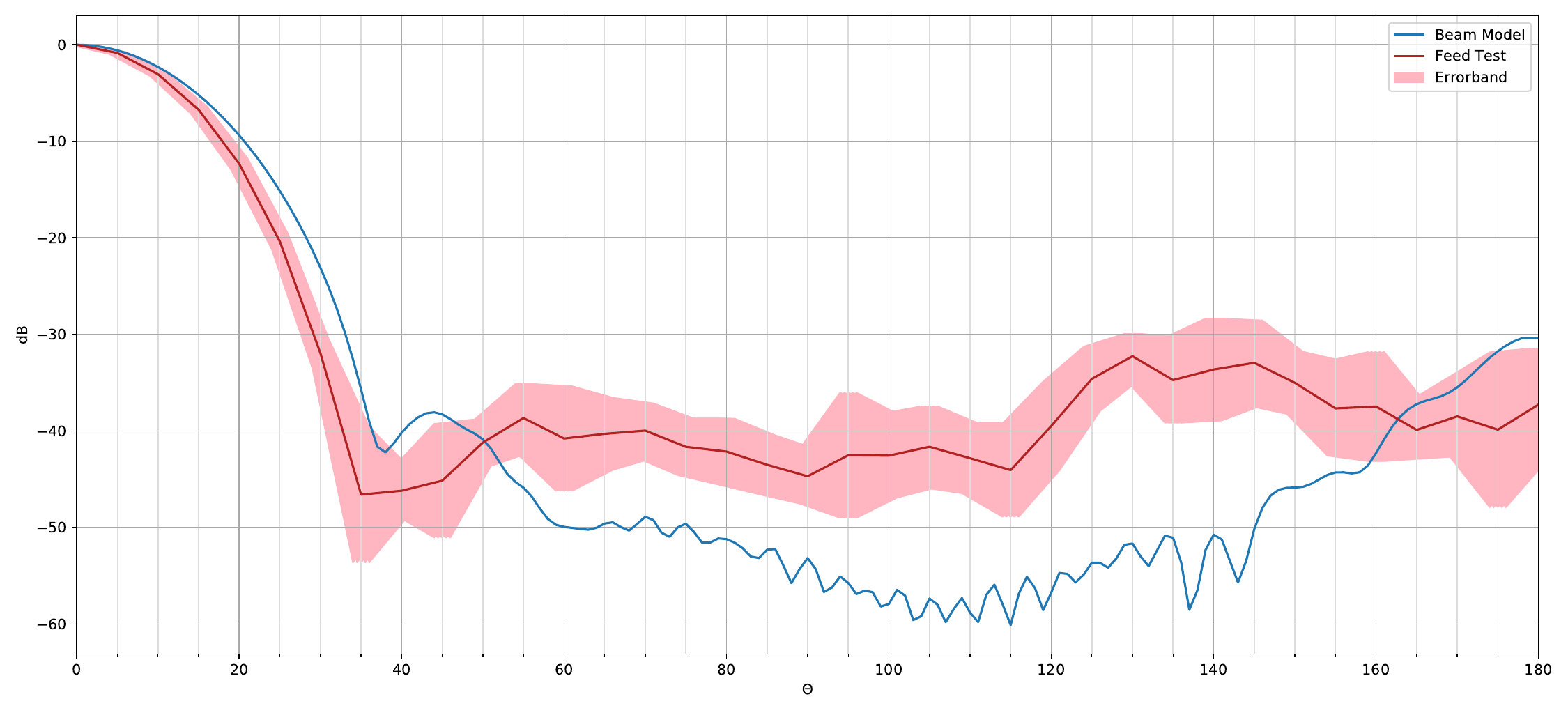}
    \caption{The beam as measured by feed tests in the field, averaged over data at 1400, 1413.5 and 1427\,MHz.  The errorband represents the standard deviation ($y$-axis) and the directional uncertainty ($x$-axis). Ground scattering of transmitter power prevented investigation of the predicted sidelobes below -40\,dB.}
    \label{simulated_corrected_beam_test}
\end{figure*}

\section{polarizer design and performance} 
\label{sec:polarizer}
At the output of each horn, a polarizer is required to convert incident waves of different hands of circular polarization into two linear outputs, which can then be fed into the receiver. In addition a transition from waveguide to N-type coaxial cable is needed. A septum polarizer is the most compact way to achieve this transformation from circular to linear. Normally a square polarizer would be used but because of the narrow bandwidth a circular design is possible and this approach was adopted since it eliminated the need for a circular to square transition. A similar design at C-band was reported by \citet{Behe1991}. A diameter of 151\,mm was chosen to maximise the distance to unwanted modes. Using the Mician Microwave Wizard, a four step septum design was optimised for return loss, isolation and relative phase between the two outputs. 
For our circular polarizer the output waveguides are semicircular so a custom waveguide to  N-type adaptor had to be modeled and optimised using the CST EM program \citep{CST_STUDIO}. These were then added to the polarizer model and the complete assembly re-optimised. The worst case prediction for the return loss was $-27$\,dB with the isolation between the two modes better than $-35$\,dB,  
and less than 1\degr\  deviation for the ideal 90\degr\  phase difference.

\begin{figure}
    \centering
    \includegraphics[width=8cm]{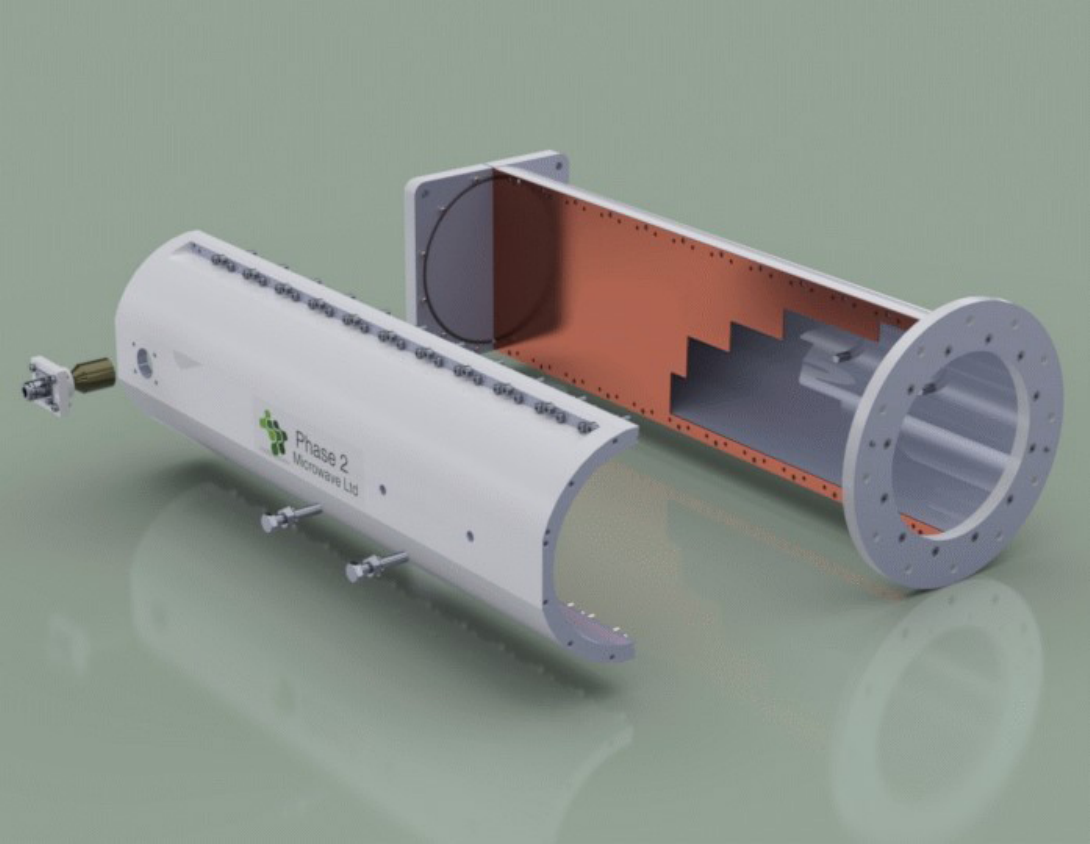}
    \caption{An exploded view of the L-BASS polarizer, illustrating the internal 4-step copper septum.}
    \label{polariser}
\end{figure}

The assembly was mechanically designed and manufactured at Phase2 Microwave Ltd (Figure~\ref{polariser}). It was realised by cutting a thick-walled aluminium tube into two halves, inserting a 2~mm thick copper septum and reassembled with a shorting plate on the connector end. All internal parts were silver plated. 

\begin{figure}
    \centering
    \includegraphics[width=8cm]{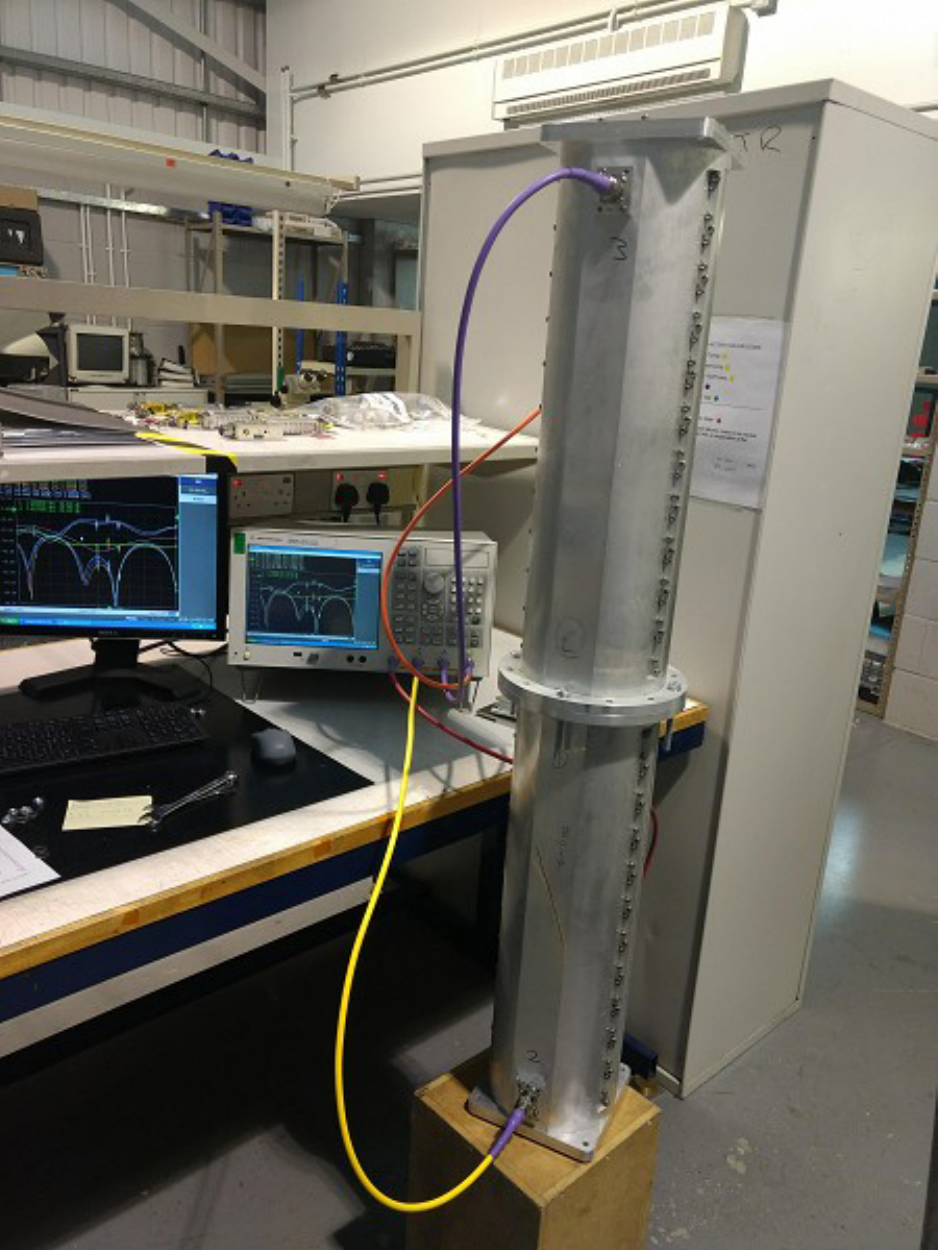}
    \caption{The two septum polarizers connected back-to-back under test in the Phase 2 Microwave lab.}
    \label{pol test}
\end{figure}

Since we had two polarizers we could test them back to back. For this a four-port Keysight E5071C vector network analyser (VNA) with a full four-port calibration proved the ideal test equipment. Figure~\ref{pol test} shows the two L-BASS polarizers under test in the Phase2 Microwave lab.

\begin{figure*}
    \centering
    \includegraphics[width=12cm]{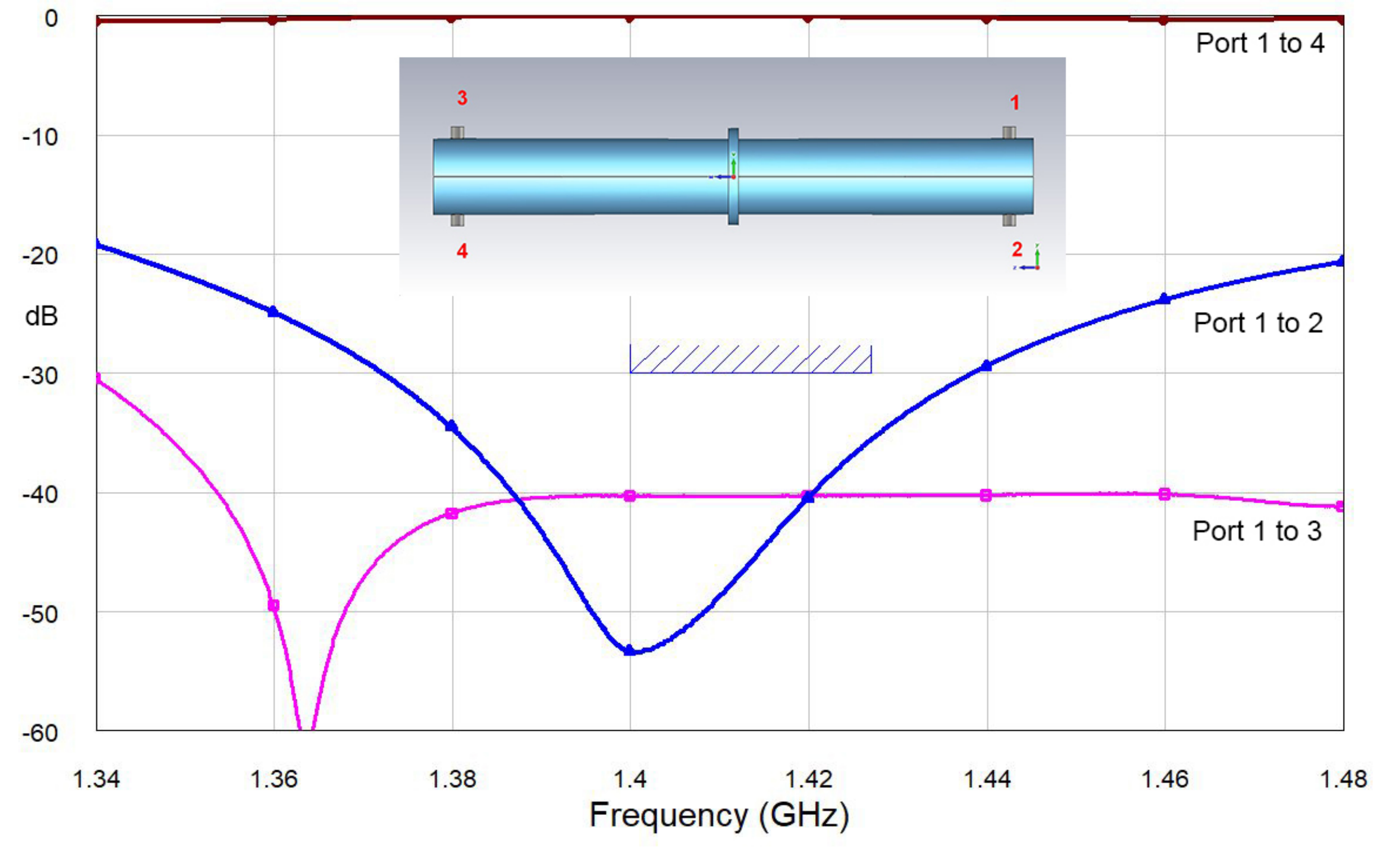}
    \caption{The results of back-to-back VNA measurements on the polarizers by Phase 2 Microwave. The hatched region shows the band of interest. The brown line (port 1 to 4; top) shows the full transmission between the two polarizers.  The blue line (port 1 to 2) shows the cross polar isolation is better than  32dB. The pink line (port 1 to 3)  shows that the cross polar isolation between the two polarizers is 40dB as predicted. }
    \label{isolation}
\end{figure*}

After small adjustments to the N-type adaptors a return loss of better than --24~dB was achieved at all ports. However, the isolation between the two polarizations  was only --28~dB not the predicted --47~dB. In order to rectify this, two tuning screws were added either side of the septum and with minimal tuning the isolation was recovered to --40~dB. Analysis using the full EM model of the back-to-back polarizers showed the effect of the tuners on the phase was negligible. VNA tests were made on the back-to-back polarizers which showed that the isolation between adjacent ports, i.e. the isolation between the two hands of circular polarization, was over --32~dB in the desired band between 1400~MHz and 1427~MHz (see Figure~\ref{isolation}). A second set of tests on individual polarizers using a shorting plate over the open end gave results which were consistent with the back-to-back measurements.

\begin{table}
\caption{Antenna insertion and return loss} 
\centering
\setlength\tabcolsep{5pt}
\begin{tabular}{l c c}
\hline 
 \textbf{Horn 1 and polarizer 1}  &  & \\
Mean insertion loss  &  & $-0.077 \pm 0.005$\,dB \\
Mean return loss & & $-31.8 \pm 0.9$\,dB \\
 \textbf{Horn 2 and polarizer 2} &  &  \\
Mean insertion loss &   & $-0.069 \pm 0.005$\,dB \\
Mean return loss & & $-27.2 \pm 1.1$\,dB \\[0.5ex] 
\hline
\end{tabular}
\label{tab:horn_pol_test}
\end{table}

\section{Antenna performance: horns plus polarizers}
\label{sec:polandhorntests}

Extensive laboratory measurements of the insertion and return losses of each horn plus polarizer combination were made with a Keysight PNA-X N5242A VNA at the Jodrell Bank Observatory. Two different configurations were used viz: i) with a shorting plate over  the aperture of a single horn mouth and ii) with the two horns placed mouth-to-mouth. For comparison we also made VNA measurements of each polarizer at JBO and at the Jodrell Bank Centre for Astrophysics (JBCA), using both shorting plate and back-to-back configurations. A full description of all the tests carried out is reported in  \citet{Zerafa_2022}. The results are summarized in Table~\ref{tab:horn_pol_test}  to which we add: 

\begin{enumerate}
\item Over the range of independent tests we were unable to detect any loss contributions from a horn+throat+flange combination, over and above that from the polarizer. Conservatively, we assess that that insertion loss of an antenna, ahead of the polarizer, must be  significantly less than 0.01~dB. 
\item The polarizer performance is consistent with that measured by Phase2 Microwave. However when averaged over all the tests polarizer 1 exhibited a slightly higher insertion loss than polarizer 2. 
\item With the horns in the mouth-to-mouth configuration the cross-polarization for a horn plus polarizer combination was found to be better than --40~dB. This is again consistent with the Phase2 Microwave measurements.
\item The overall performance of the polarizers falls off towards the top end of the observing  band (1427~MHz). For this reason science observations will be restricted the observing band in use to 1400--1425\,MHz. 

\end{enumerate}

Knowing the beam properties on the sky is vital. The simulations used in the design process for the horns did not include the effects of the polarisers so new simulations have been performed taking into account their effect. The results are plotted in Figure~\ref{beam_vs_solar}. The predicted FWHM beamwidth in the middle of the band is 23\degr\.  and the sidelobes are generally $\leq$-40 dB. It should be noted that different software packages can give different results for the strength of the back-lobe. Multiple solar transit observations (also shown in Figure \ref{beam_vs_solar}) have independently confirmed the predicted FWHM (measured to be $23\pm 0.25$\degr) and verified the shape of the main lobe down to $-25$\,dB from the peak. The main lobes for the East and West horns are identical within the errors. However, the sun is not strong enough for us to explore the far-out sidelobe levels so before full assembly of the telescope one of the horn/polarizer systems was tested in a purpose-built feed test facility located in a field at Jodrell Bank.  The horn was mounted on a turntable and a linearly polarized CW signal was radiated towards the antenna under test and the power received as the feed was rotated in steps was measured using a spectrum analyser. The results of this test are shown in Figure \ref{simulated_corrected_beam_test}.  Our ability to confirm the sidelobe structure as shown in Figure \ref{simulated_corrected_beam_test} is limited by scattering but the measured far-out sidelobes are typically at the $\leq-40$ dB level. They do not confirm the $-30$\,dB backlobe in the model but as mentioned earlier different EM software packages predict different patterns of the ultra low-level sidelobes. We note that the main beam measured in the field is significantly narrower than the prediction and that measured on the Sun.
The field measurements involve geometric corrections based on the distance of
the positions of the phase centres of both the transmitting antennas and 
beam pattern of the transmitting antenna. Uncertainties in the assumed values
for any or all of these could account for the discrepancy. However, these uncertainties
do not significantly affect the determination of the sidelobe levels which was the main reason for making field measurements.
In practice the sidelobe pattern is affected by mounting the antennas on the L-BASS structure. However the ground screen ensures that the sidelobes end up pointing at the cold sky.\\ 

The main lobe can be approximated by a Gaussian and has FWHM of $\sim23$\,degrees so its solid angle is 1.13 (FWHM)$^2 \approx 600 {\rm \, deg}^2$; the sidelobes therefore subtend a total of $41250 - 600 \approx 40,650$\,deg$^2$.  If we take their mean level to be $-40$\,dB from the peak (which is conservative on the basis of the beam modelling) then their effective solid angle is $\sim 4.1$\,deg$^2$ i.e. $\sim 0.7$\% of that of the main lobe. Because L-BASS is surrounded by a tall ground screen with a metal floor, virtually all the sidelobes are directed onto the cold sky. Measurements of the level of residual ground pick-up will be reported in Paper III.

\section{4-m cable performance} 
\label{sec:4-m cables}

The antennas are connected to the receiver with Gigatronix LBC-400 coaxial cables 4.04\,m in length. These are critical components since any variations in their transmission characteristics, due to changes in ambient temperature and/or flexure as the antennas are moved in elevation, will affect the power levels entering the receiver and mimic a change in sky brightness temperature. For these reasons extensive tests have been performed on these cables. 

\subsection{\bf Cable losses and temperature susceptibilities} 
\begin{figure}    
\centering
\includegraphics[width=8cm]{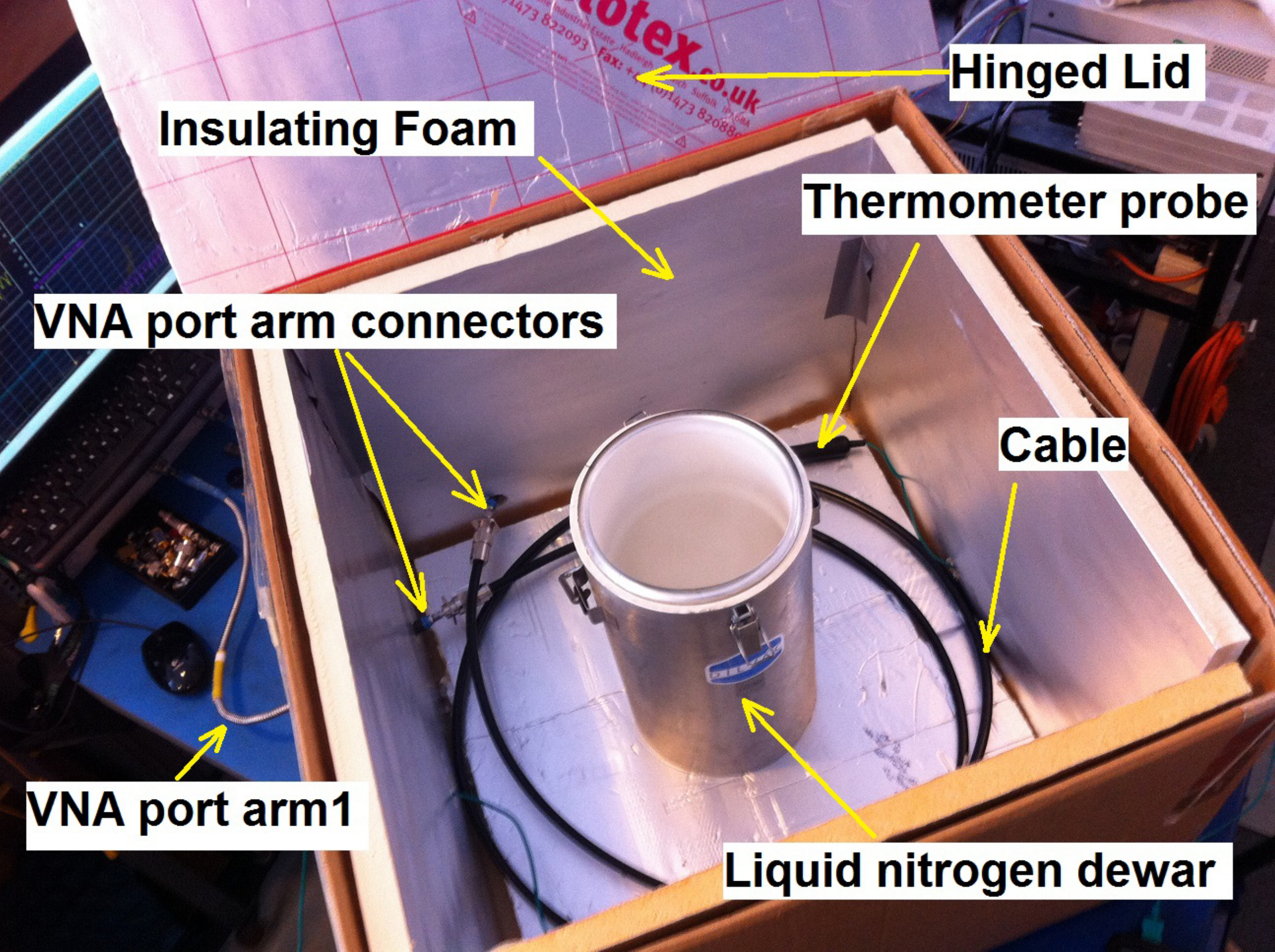}
\caption{Experimental apparatus set-up for liquid nitrogen cooling of coaxial cables.}
\label{fig:insulationbox}
\end{figure}

The ohmic losses in the two 4.04-m LBC-400 cables with their connectors, 0.740~$\pm 0.005$\,dB at 24\,\degr C,  are very similar with the connectors contributing 0.041 dB to this total. The variation in loss in the combination of cable plus connectors was measured over a wide range of temperatures by loosely coiling a 3-m length inside a foam insulated box and connecting it to a VNA (see Figure~\ref{fig:insulationbox}).
A Dewar flask of liquid nitrogen at 77\,K was placed centrally within the box and the cable was cooled by blowing room temperature air over the dewar, evaporating the liquid into  cold gas which then cooled the cable under test to a desired starting temperature. The cable was then allowed to warm up to ambient laboratory temperature and measurements of the insertion loss were taken during this process. The loss is linear (in dB) over the range $-20\,\degr$C to $+24\,\degr$C and, scaled to a 4.04~m length, corresponds to a temperature susceptibility of 1.4~$\pm~0.05~$mdB/\degr C for the cable alone. From separate tests we estimate that the N-type connectors themselves have a temperature susceptibility of $\approx 0.4$\,mdB/\degr C. We separated the contributions of cable and connectors anticipating that the connectors would be at significantly different temperatures from the cable; warmer in the receiver box and cooler at the polarizer. In practice, the cables are: i) thermally insulated and actively heated with a low voltage wire coiled along their lengths which reduces the temperature gradients; ii) fitted with 7 temperature sensors with two additional sensors placed as near as physically possible to the connector ports (see Section \ref{temp_monitor}).

The LBC 400 cables connect the antenna to the inputs of the first magic-tee hybrid.  The outputs of this hybrid are connected to the first LNAs via equal (1~m) lengths of semi-rigid SMA cable. Loss measurements of an SMA cable in a heated box revealed a temperature susceptibility of 1\,$\pm~0.06$~mdB/\degr C/m. 

The temperature profile data and the measured temperature susceptibilities enable the cable losses and associated radiometric noise emission to be modelled and corrections made to the system performance. This is part of the "passives model" used in the calibration process and more details will be given in Black et al. (in prep) hereafter Paper III.

\subsection{\bf Cable flexure} In operation, as the horns move in elevation,  the cables connecting the horns to the fixed receiver flex and any changes in transmission characteristics as this happens will also mimic a real change in sky brightness temperature. We therefore made careful laboratory measurements of cable loss investigating flexures typical of those encountered during science observations. The results showed that the loss does not change by more than 0.001~dB as the cable is lightly flexed, provided the first 0.25~m of the cable close to each N-type connector is held absolutely rigid, thus preventing  mechanical strain creating transmission changes. We therefore constructed robust cable harnesses both at the polarizer end (see Figure~\ref{horn-and-polariser}) and where the cables are attached to the receiver inputs.

\subsection{\bf Cable connection reproducibility.} In the first phase of the project, during which the sky brightness temperature will be compared to that of the NCP, the 4.04~m cables connections to the polarizers do not need to be touched. However, in the second phase of the project, in which the absolute brightness temperature of the NCP is to be established, the output from one of the antennas will be replaced with the output from a cryogenically cooled reference load whose radiometric temperature is close to that from the NCP antenna. This will require disconnecting one of the cables at the polarizer end and connecting it to the output of the cryogenic load. This process will inevitably introduce small but unpredictable changes in propagation properties.  Since the signal out of a polarizer and/or the cryogenic load will have a radiometric temperature of $\sim 10$\,K, an unpredictable change in the transmission coefficient of 0.001\,dB after the disconnect/connect is equivalent to an unpredictable change in the radiometric input to the first hybrid of 0.0023\,K; this is well below our goal of 0.1\,K absolute accuracy. However, more importantly, a change of ohmic loss of 0.001\,dB will add or subtract a radiometric contribution of $\approx 0.067$\,K at the typical physical temperature of the polarizer port ($\approx 290$\,K); this would be a significant contribution to the error budget.

Ensuring disconnect/reconnect reproducibility at the 0.001\,dB level presents a significant technical challenge and may well be the single most important factor limiting the accuracy of our final absolutely calibrated map. Because of the importance of this issue extensive disconnect/reconnect tests have been performed on a range of different connector types. 
At each end the 4~m cables are terminated with type-N connectors. Our lab tests show that when new type-N junctions are disconnected/reconnected the changes in measured insertion loss have a typical spread of $\sim 0.003$\,dB which is not good enough to meet our 0.1~K goal. Therefore, between the polarizer output and the cable connector we have  inserted genderless coaxial APC-7 (Amphenol Precision Connector 7~mm) type connectors. These are designed for use with VNA calibration kits and metrology and are designed to maintain a reflection coefficient of 0.001\,dB repeatably on disconnect/reconnect. When tested with our best VNA the insertion loss values are repeatably close to the desired 0.001\,dB level. 

\section{Receiver design and performance} 
\label{sec:receiver}

\begin{figure}
\centering
\includegraphics[width=8cm]{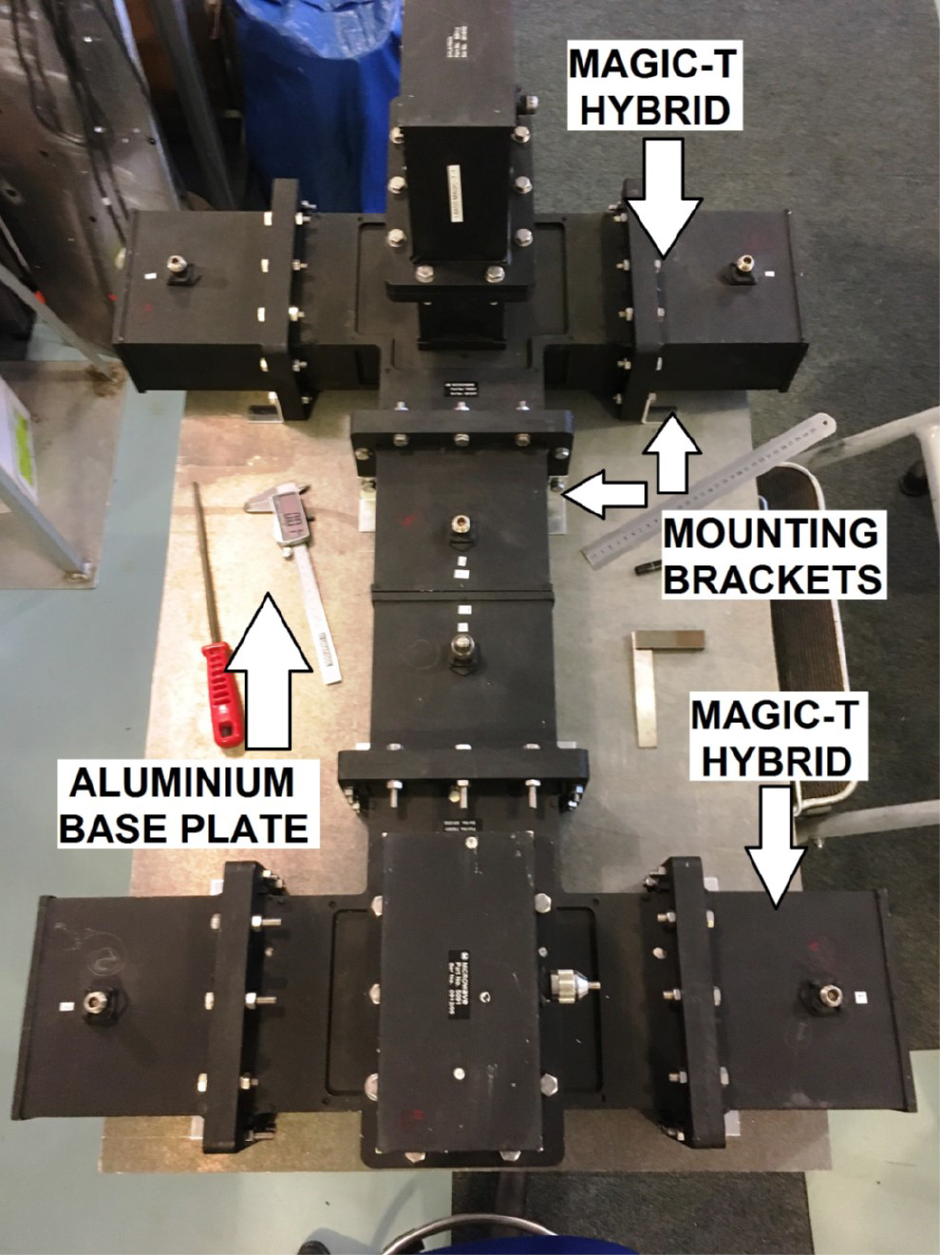}
\caption{The hybrids mounted on the aluminium base.} 
\label{Magic Ts} 
\end{figure}

\begin{figure}
\centering
\includegraphics[width=8cm]{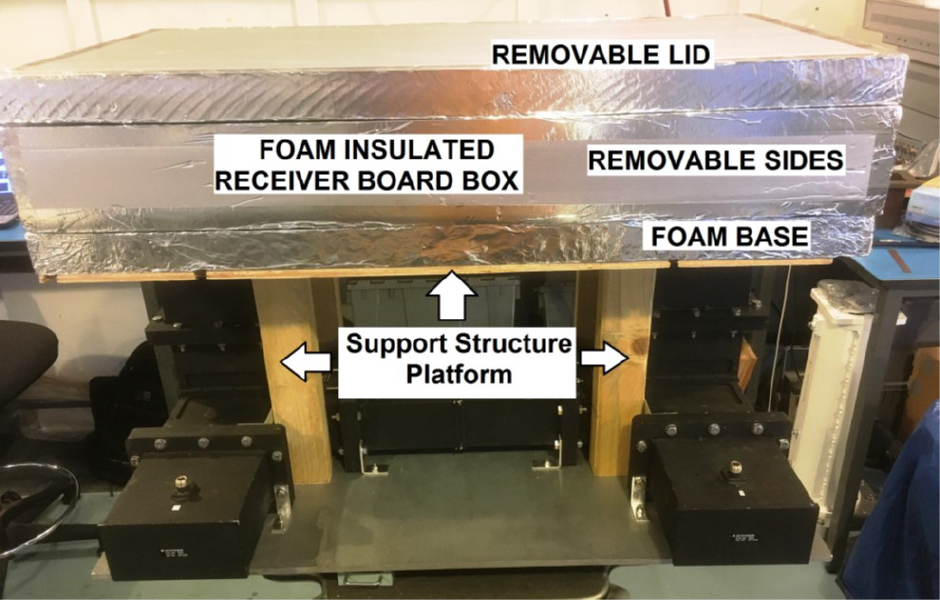}
\caption{The receiver is located within an insulated box supported above the hybrids.} 
\label{mounted_box} 
\end{figure}

\begin{figure}   
\centering
\includegraphics[width=8cm]{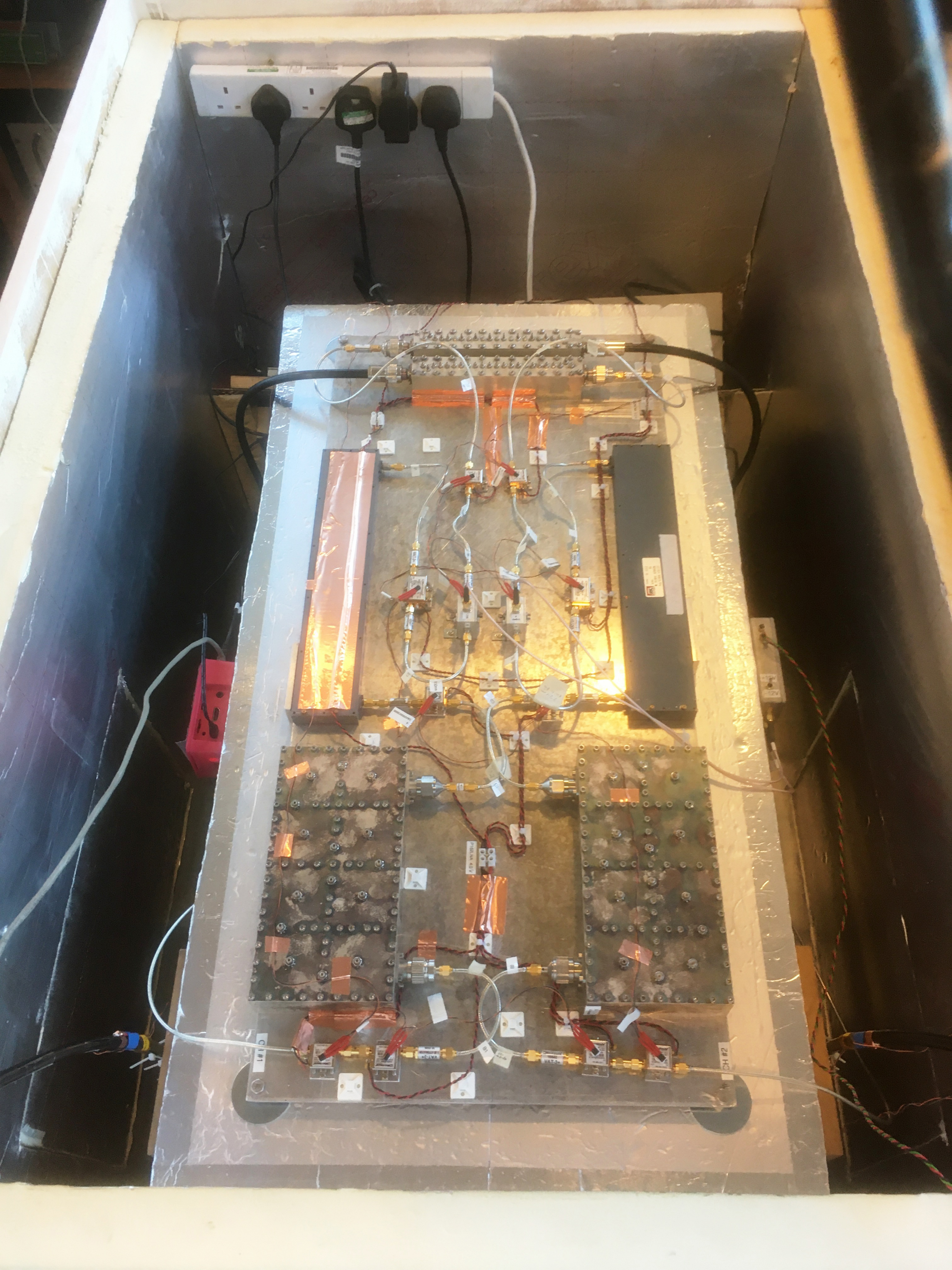}
\caption{The active receiver is housed within an inner insulated box; here both inner and outer box insulated lids have been removed.} 
\label{receiver_in_box} 
\end{figure}

In common with WMAP and {\it Planck} LFI, the L-BASS receiver uses a differential continuous comparison architecture. 
A pair of magic-tee hybrids (hereafter 'hybrid'), fabricated in waveguide size WG6\footnote{also known as WR650 or R14.} (Fig.~\ref{Magic Ts}) are used to split and then recombine the signals: the input and output ports of the receiver as a whole are the inputs of the first hybrid and the outputs of the second. The active receiver is housed in an insulated box held above the hybrids by means of a wooden support structure (Fig.~\ref{mounted_box}). This combination is housed within a larger insulated box (Fig.~\ref{receiver_in_box}) which is mounted on the main telescope structure between the antennas. The air inside this outer box is heated to above ambient temperature\footnote{$30\,^{\circ}$C in winter and $40\,^{\circ}$C in summer.} and thermostatically controlled to better than $1^{\circ}$C. During calm nights the temperatures of the components within the inner box are stable to $\sim 0.1^{\circ}$C.

\begin{figure*}   
\centering
\includegraphics[scale=0.6]{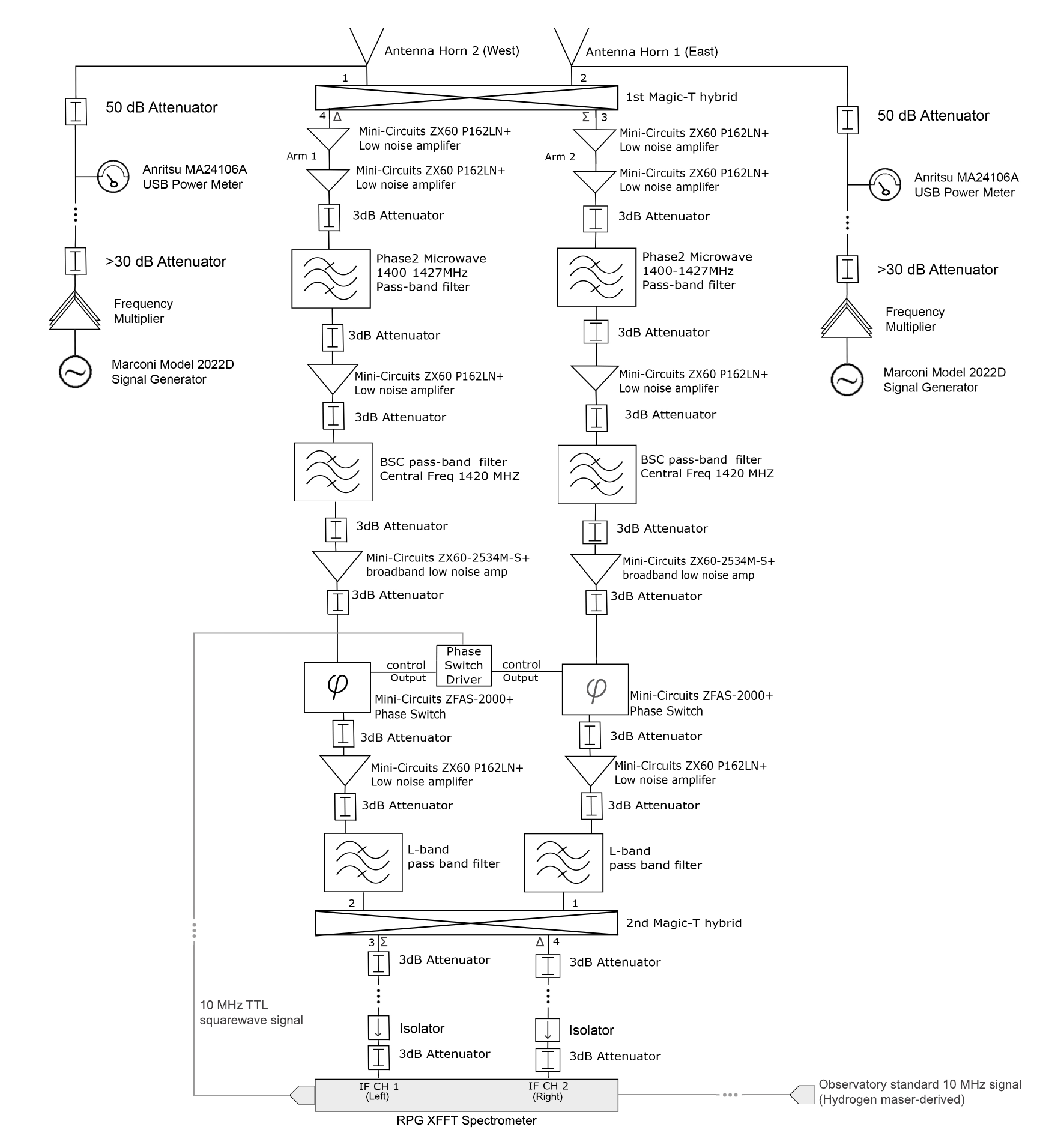}
\caption{L-BASS receiver schematic diagram, showing the two component chains comprising LNAs, attenuators, pass band filters and phase switches positioned between the first and second hybrids.   The outputs from the second hybrid are fed via 55-m cables into the RPG-XFFTS digital spectrometer located in an adjacent building. The CW signal injection system into each horn is described in section \ref{CW system}}
\label{fig:MiniCircuits-schematic} 
\end{figure*}

A component-level block diagram of the receiver is shown in Fig.~\ref{fig:MiniCircuits-schematic} and the layout of the actual components shown in Fig.\ref{receiver_in_box}.
The front-end RF gain is provided by two independent chains of low noise amplifiers (LNAs)
between the hybrids.  In each chain there are two LNAs followed by the band-defining filter (1400--1427\,MHz) manufactured by Phase2 Microwave. These filters protect the receiver from unwanted signals from DC up to 4.2 GHz; they provide >~40~dB of rejection only 4~MHz away from the band edges and the rejection is below the 95~dB noise floor from 0 to 1.3\,GHz and from 
just below 1.6\,Ghz  to 4.2\,GHz.  Above 4.3\,GHz the rejection is not 
controlled.  There are two further bandpass filters and three more RF amplifiers in each chain. The additional filters are there to suppress out-of-band signals which can be aliased into the band of interest in the RPG-XFFTS digital spectrometer, which is operated in the 4th Nyquist zone (see Section \ref{spectrometer}). The additional filters not only provide protection against very strong out-of-band signals (in particular mobile phone base stations in close proximity to the observatory) but also the 'pedestal' of noise power which is generated by broad band amplification after the initial bandpass defining filters. In order to minimize potential oscillations in the amplification chains which can arise from unwanted reflections, 3\,dB attenuators are inserted between many of the components, symmetrically in each chain. To the same end 0.1\,$\mu$F capacitors were soldered across the bias pin turrets of the LNAs and in some places the d.c. power supply connecting wires to each LNA were held down to the metal base plate with copper tape. This also helped in the mitigation of cross-talk within the receiver.

In each chain there is also a phase switch, one of which is driven by a 10\,Hz square wave produced by the spectrometer. The driven switch changes the phase of the signal by 0\degr\ or $\approx 180\degr$; the other switch is there to maintain amplitude and phase symmetry between the chains\footnote{The phase shift produced by the Mini-Circuits ZFAS-2000+ unit is $\sim$175.5\degr in 1400~MHz to 1427~MHz band; we quantify the effect of the shift not being exactly $180\degr$ in Paper III.}.
The 55 m cables connecting the receiver to the spectrometer are part of the post-amplification signal path and therefore contribute a negligible amount of noise to the system. Although buried in a cable trench, their loss will vary with temperature seasonally, with a peak summer/winter variation of <~10\%, and with diurnal variations of a few percent. However, with the combination of phase switching, double differencing and continuous gain calibration (see section \ref{CW system}), the receiver arrangement enables the effects of variations in gain or loss after the second hybrid (including the 55~m cables) to be eliminated, provided that the timescale of variation is significantly longer than (10\,Hz)$^{-1}~=~0.1$\,s.

\subsection{Receiver tests and system monitoring}
Extensive testing of the receiver and its components has been undertaken in the knowledge that its performance is optimised if the two chains have closely similar performance and the isolation between them is maximised. The following measurements were performed on different aspects of the system.

\begin{enumerate}
\item{\bf Individual components~}
All the amplifiers, filters and cables were individually measured with a VNA and all found to meet the manufacturer's specifications. For the hybrids we have  the manufacturer's measurements of the S-parameters. The overall gain of each amplification chain between the hybrids was measured in the system passband: for amplifier chain arm 1 the gain is 90.89\,dB and for amplifier chain arm 2 the gain is 90.95\,dB, thus they are matched to within 0.06\,dB. This is a good level of performance in consideration of the $\approx 118$\,dB gain the amplifiers produce in each chain arm before additional attenuation.  There are two types of Mini-Circuits amplifierss; the ZX60 P162LN+ LNAs have a lab measured noise figure of $\sim 0.66$\,dB and the ZX60-2534M power amps have a quoted noise figure of $\sim 2.94$\,dB.  The noise temperatures of the active receiver chains are set by the pair of ZX60-P162LN+ LNAs at the input to each chain, since their combined gain exceeds 38\,dB. The amplifier chain noise temperature susceptibility to physical temperature changes is about +0.26~K/\degr C. At typical L-BASS physical operating temperatures we estimate the receiver noise temperature $T_{\rm Rx}$ contribution to the system temperature $T_{\rm sys}$ is $\approx 52$\,K\footnote{Together with $\approx 83$\,K contributed from passive components in the pre-amplification signal path this gives a system temperature $\approx 135$\,K.}. The gain of each individual LNA is $\approx  19.5$\,dB, and the gain of each power amp is $\approx 39.5$\,dB.

\item  {\bf Isolation~} For optimum performance the cross-talk between the chains should be as low as possible i.e. with no phase switching a signal entering one of the inputs of the first hybrid should only appear at one of the outputs of the second hybrid. In order to achieve this isolation, the phase delays in the two chains between the hybrids must be equalised. After the receiver had been assembled we had no \textit{a priori} knowledge of the precise differential delay between the two multi-component chains.  We therefore proceeded empirically by inserting a series of RF cables of different lengths into one chain and measuring the changes in power at the outputs of the second hybrid  when a noise diode, connected to input 1 of the first hybrid, was switched on and off; a matched $50\,\Omega$ load was connected to input 2. Figure~\ref{fig:receiver_crosstalk} shows the changes in the output power levels as the path length in chain 1 was varied. As theory predicts \citep{Jarosik_2003} the variations are cosinusoidal. It was found that the signal is maximised in output 4 of the second hybrid (see Fig.~\ref{fig:MiniCircuits-schematic}), and the leakage of signal into output 3 is a minimum, for a cable path difference of  $-74\pm 1$\,mm; this corresponds to an electrical path difference of $106.5\pm 1.4$\,mm, or $181\pm 2$ degrees of phase, at the band centre, 1413.5\,MHz. For this path difference the power from output 4 changes by $50.700\,\mu$W whilst from output 3 it changes by $0.039\,\mu$W; the isolation ratio is therefore $\sim1300$:1. As the path difference is reduced to zero the power transferred swaps over between the outputs; the change in output 3 is $49.202\,\mu$W and in output 4 it is $0.039\,\mu$W, an isolation ratio of $\sim 1260$:1 These results show that the chains were actually close to phase balanced as constructed but in practice the 74~mm cable path difference was retained in chain 1\footnote{for simplicity this is not shown in Fig. 12}.  The fact that an isolation of $>$30\,dB can be achieved confirms that, across our $\sim 2\%$ observing band, the hybrids approach their idealised phase performance.

\begin{figure} 
\centering
\includegraphics[scale=0.48]{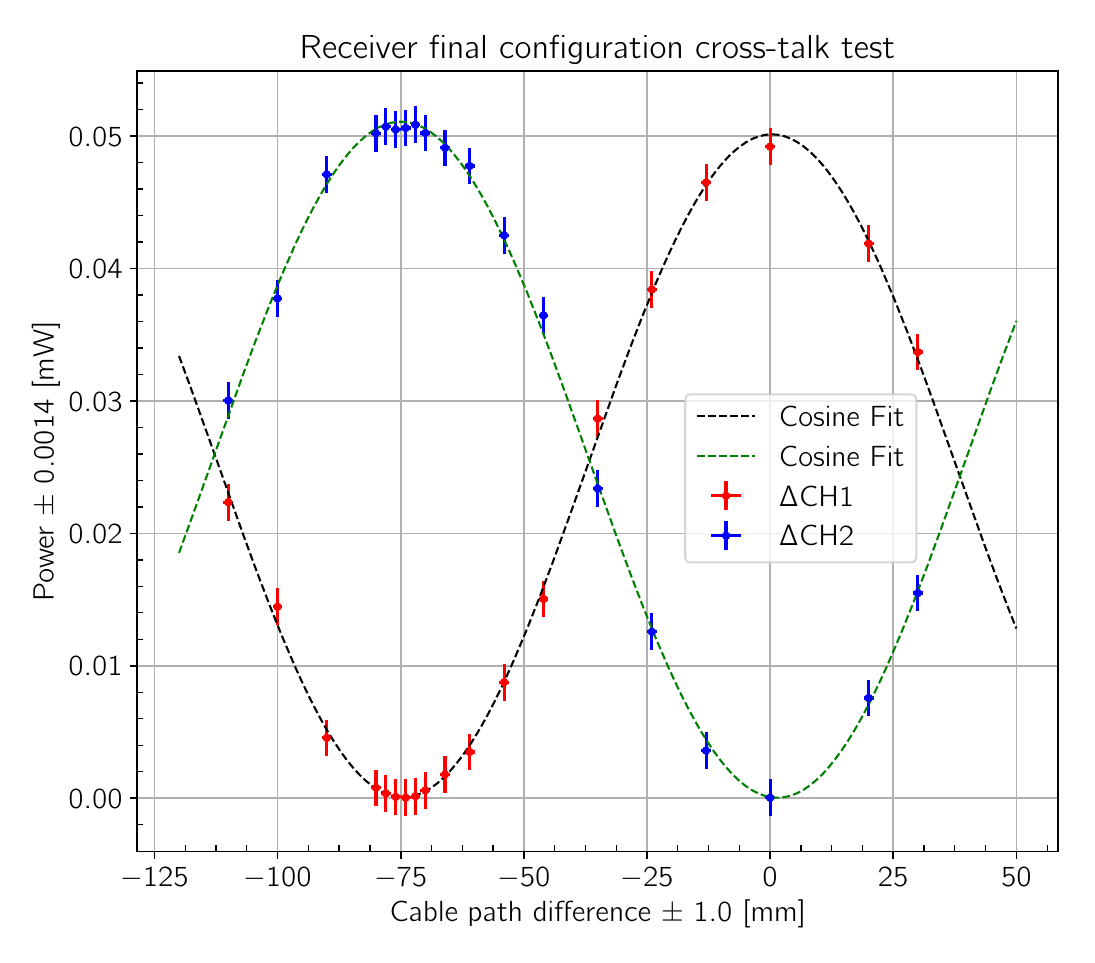}
\caption{Receiver cross-talk (isolation) test. The power changes observed in the two outputs of the second hybrid when a noise diode signal is applied to an input of the first hybrid and additional cable lengths are inserted in one of the chains between the hybrids. The red data show the power changes seen in output 3 (connected to input CH1 of the RPG-XFFTS digital spectrometer)  and the blue data show the power changes in output 4 (connected to input CH2 of the RPG-XFFTS). As the differential path is varied the maximum power change swaps over between the two outputs (see text for further discussion).}  

\label{fig:receiver_crosstalk}
\end{figure} 
\item {\bf Receiver chain temperature susceptibility} The characteristics of the components in the receiver chains, particularly the gains and noise temperatures of the LNAs, will change as their physical temperature changes. Prior to installation on the telescope we therefore tested the receiver's temperature susceptibility in the lab. We placed the receiver inside a temporary foam enclosure and heated the interior air temperature to 48\,\degr C, well above of the ambient laboratory temperature.  The output from a receiver chain was monitored with its input terminated with a matched load situated outside the box and equilibrated to the ambient temperature of the laboratory. The interior heating was then switched off and the chain allowed to cool down to 25\,\degr C. The results shown in Fig.~\ref{fig:temp_suscp} shows that the relationship between the total power output from the chain, when measured in dBm, and its physical temperature is linear. The proportionality is --0.0595 $\pm 0.0003$~dB/\degr C.  The results from the two chains are very similar.  On the telescope the receiver chains are housed inside a thermally insulated environment in order to minimise temperature effects on the receiver (see Fig.~\ref{receiver_in_box}). Within the inner box the physical temperatures change by only $\sim 0.1\,\degr$C on calm nights. Total power variations in both active chains will therefore be limited to $< 0.15\%$ and the difference between them will be $\ll0.1\%$. 

\begin{figure}
\centering
\includegraphics[scale=0.45]{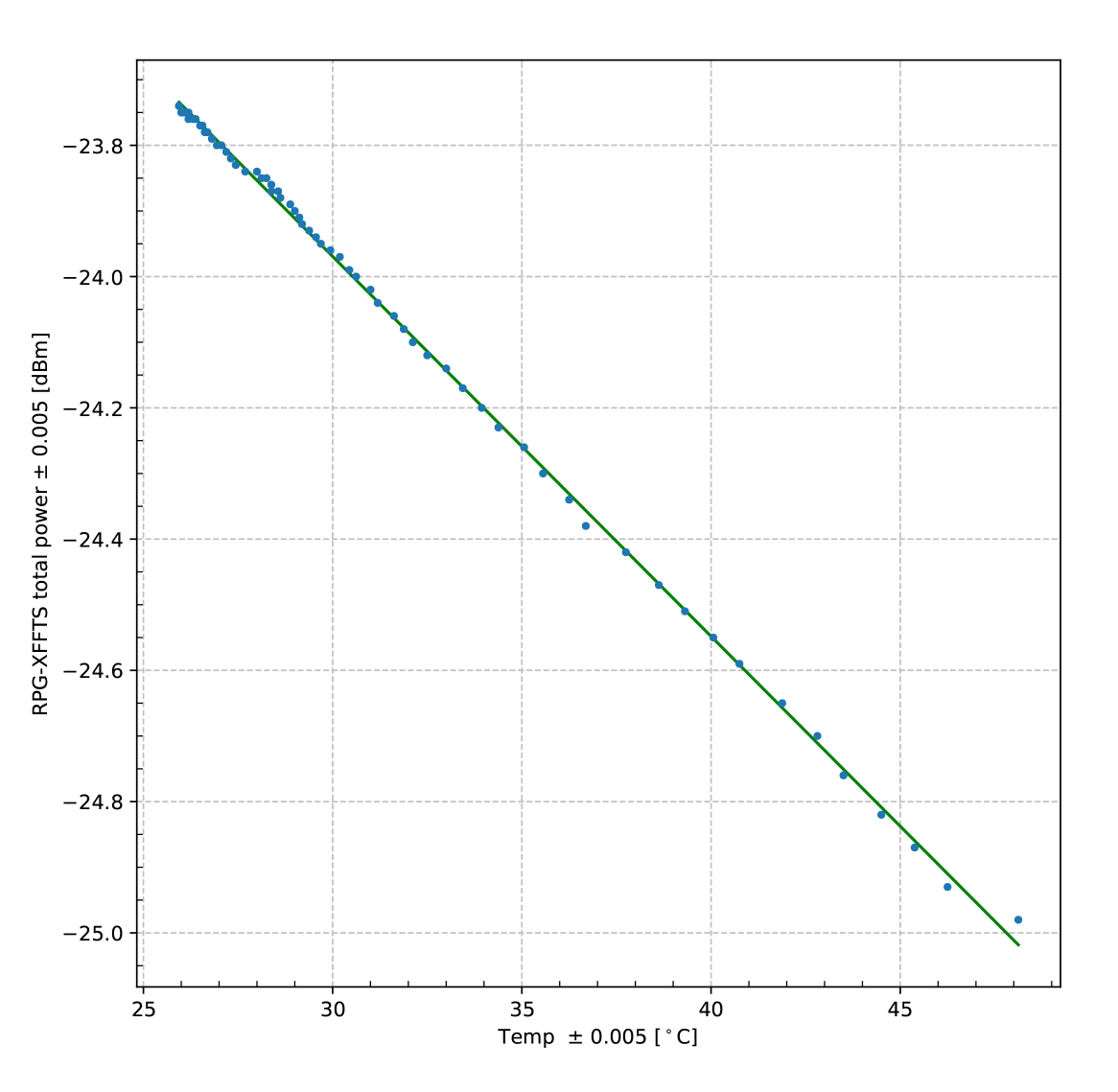}
\caption{The total power output of the receiver chain channel 2 as a function of physical temperature for a sample Mini-Circuits P162 LNA within it; the input is a constant temperature matched load. The power output measured in dBm is linear over the range 25\,\degr C to 48\,\degr C and the temperature susceptibility is $-0.0595~\pm~0.0003$~dB/\degr C.}
\label{fig:temp_suscp}
\end{figure}

\item {\bf Receiver linearity.} The linearity of the receiver chains as a function of input signal level was tested in two ways; first by injecting a CW signal at 1413.5~GHz into each of the receiver chains using an Agilent 83565L signal generator.  A high degree of attenuation was inserted in the signal generator output and this was reduced in steps; the resultant output signal was recorded with an Anritsu MS2720T spectrum analyser. The results are plotted in Fig.~\ref{fig:receiver_chains_linearity} which shows that both receiver chains respond linearly to changes in input power level over a wide range. This behaviour was cross-checked by injecting a broad-band signal from an Agilent 346B noise diode into chain 2 and varying the input attenuation; the output power was measured with an Agilent ML2438A power meter. The linear gain response was confirmed. All these tests were carried out in the stable laboratory temperature environment. 
\end{enumerate}

\begin{figure} 
\centering
\includegraphics[scale=0.36]{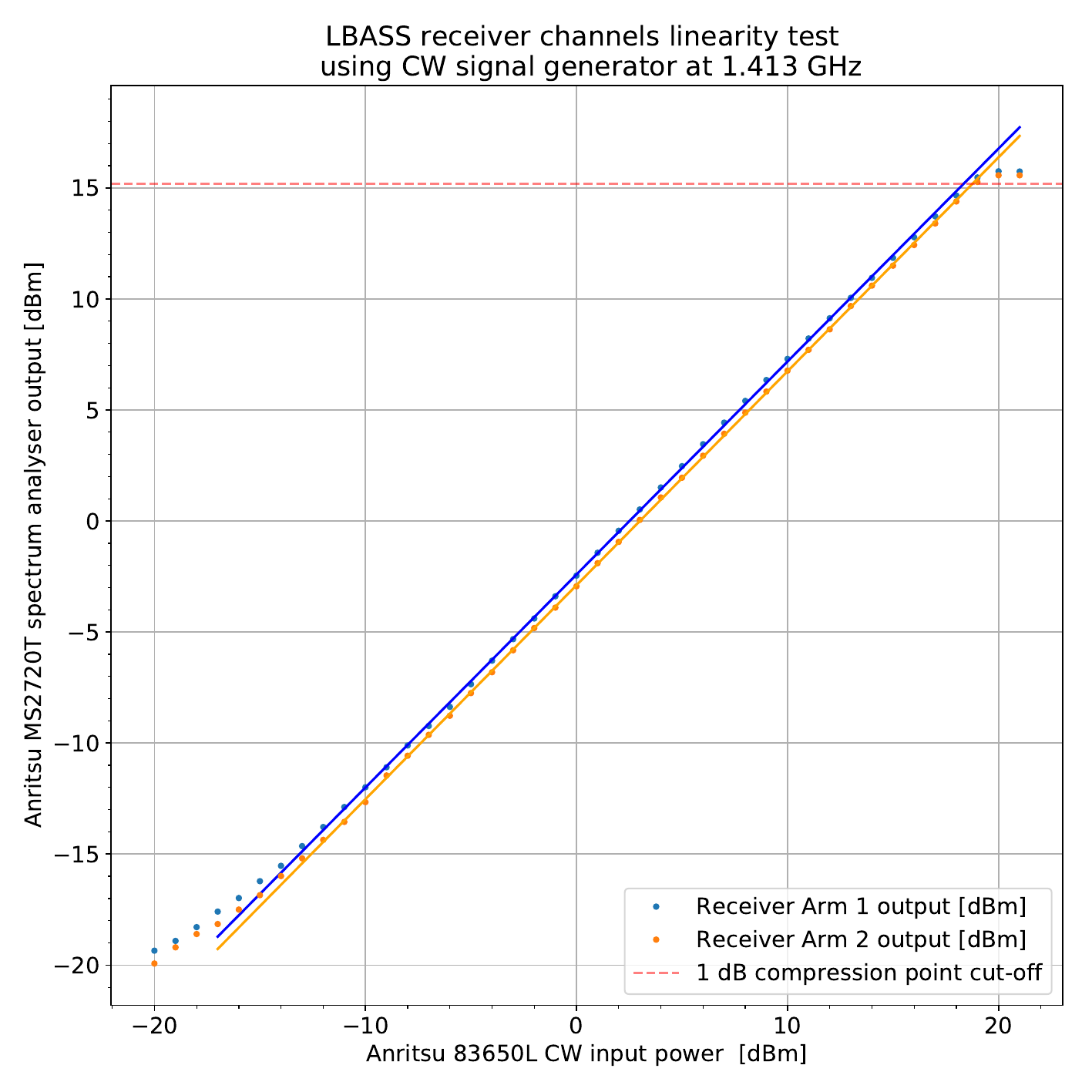}
\caption{The gain linearity of the receiver chains was measured  by injecting a CW signal at 1413.5\,MHz and varying the input attenuation.   We see as the CW signal increases in power from the noise floor (without attenuation it is $-90$\,dB), become linear in its gain until it reaches the 1\,dB compression point indicated by the red dotted line. }
\label{fig:receiver_chains_linearity}
\end{figure} 

\subsection{The Digital Spectrometer}
\label{spectrometer}
The two output signals from the second hybrid are fed through cables into a digital backend, the \textit{eXtended Fast Fourier Transform  Spectrometer} (XFFTS) manufactured by Radiometer Physics GmbH (RPG). As shown in Fig \ref{fig:MiniCircuits-schematic} there are attenuators and isolators in the signal path between the second hybrid and the inputs to the spectrometer; these suppress any standing waves within the cables.\\ The RPG-XFFTS is housed in a metal-lined, temperature-controlled room (referred to as the screened room) in a building $\sim$40~m away from the antennas and front-end receiver; the spectrometer is frequency locked to the Observatory hydrogen maser-derived 10\,MHz frequency  standard. The room is temperature controlled to better than 1\,$\degr$C and the unit itself is surrounded by a insulating foam box so as to keep the temperature of the ADC amplifiers as stable as possible hence to minimise gain fluctuations prior to the signal digitization. 

The spectrometer is operated in the 4th Nyquist zone and produces spectra of the two input signals covering a 450\,MHz bandwidth with each of the 8192 frequency channels having a width 54.932 kHz; across the utilised 1400 - 1425\,MHz band there are, therefore, 456 frequency channels.  The two spectra produced by the spectrometer are fed into a PC for post processing and data storage. The full 450\,MHz bandwidth can be displayed in real time via the spectrometer's LABVIEW interface \citep{labview}, but only 718 channels are saved, centred on the bandpass but extending far enough out to include channels with negligible front-end gain (see Sec.~\ref{sec:data_logging}). The spectrometer also produces a TTL square wave output which is used to drive one of the phase switches in step with the data sampling.

\subsubsection{RPG-XFFTS linearity}

The RPG-XFFTS was not designed to operate as a broad-band power meter and so it was necessary to check its linearity at the power levels typical of L-BASS science observations.  To do this a noise signal from Noisecom NC1110A high power noise diode was first injected  into a calibrated Anritsu ML2438A power meter through a band-pass filter centred on 1420 MHz and followed by a series of 3dB attenuators, and then the same signal injected into the RPG-XFFTS where the signal strength was measured with its monitor software. This process was repeated over a wide range of attenuations. The comparison showed that RPG-XFFTS exhibited a high degree of linearity over input power levels from -40 dBm to -11 dBm.  During scientific observations the input power levels will range from -20 to -23 dBm with the highest values occurring during calibration observations on the Sun. For the great majority of the time the input power levels will vary by less than 1 dBm. 

\subsubsection{RPG-XFFTS Temperature dependent gains}

The spectromter generates a square wave Transitor-Transitor Logic (TTL) signal at 10\,Hz which, via a phase switch driver in the receiver chains, acts to repeatedly change the phase switch between 0 and $\pi$ states.  This effectively alternates the signals at the output ports of the second hybrid, ensuring that the signal associated with one antenna passes through both 55\,m cables. There are two inputs to the spectrometer, designated Left (L) and Right (R), and two receiver phase-switch states (0 and $\pi$), giving us 4 outputs: L,0 and R,$\pi$ associated with the East horn (E), and, L,$\pi$ and R,0 associate with the West horn (W).\\
 The L and R inputs to the spectrometer have independent gain factors and these gains change in relation to the operating temperature of the ADC.  If the temperature dependences are different this could alter the R/L gain ratio and thus the ratio of powers recorded in the two phase-switch states from the same horn.  
 Fortunately, the temperature dependences are similar and  any issue can be easily mitigated by diligent control of the RPG-XFFTS operating temperature.  The normal temperature range for the ADC during typical L-BASS operations varies by less than $\pm 1 \,$C over periods of several hours.  Over such ranges the ratio between the two temperature dependent gain factors is 0.975 for the East horn signal (R,$\pi$/L,0) and 0.943 for the West horn signal (R,0/L$\pi$) and both remain stable at $\pm 0.1\%$.  
 \\
 As in WMAP \citep{Jarosik_2003} and Planck-LFI \citep{Seiffert2002}, L-BASS utilises a scheme of `double differencing' which serves to further mitigate back-end gain instabilities.  The complete formalisation of our scheme of differencing will be provided in Paper III.

\subsection{The gain monitoring system}
\label{CW system}

Changes in output power of the system as a whole will arise from a combination of changes in the gains (multiplicative) and in noise temperatures (additive) of the LNAs but also in losses and thermal emission from the various connecting 
cables. As mentioned in Paper I to keep track of the system gain we  inject well-calibrated CW (continuous wave) signals into the horns in the  manner suggested by  \citet{Pollak_2019}. In our case the CW is injected into a sky channel via the spare polarizer port relying on the $\approx -30$~dB port-to-port cross-coupling. The general arrangement is shown towards the top of Figure \ref{fig:MiniCircuits-schematic}. 

The CW signals are produced by Marconi Model 2022D signal generators locked to the Observatory maser-derived 10\,MHz frequency standard and then quadrupled in frequency before being transferred to the telescope via an independent pair of 55m cables. The CW frequencies are different for each horn and are chosen so that they fall in the centres of two of the 55-kHz spectrometer channels. On arrival at each antenna the CW signals are split with half being continuously monitored with Anritsu Model MA24106A USB power meters. The other half of the CW signal is attenuated by 50~dB before injection into the spare polarizer port. By comparing the power meter readings with the strength recorded after passage through the entire receiver a continuous measure of gain changes is obtained.  

The splitters, the MA24106A power meters and the 50~dB attenuators are housed in insulated boxes and mounted on substantial metal plates to provide thermal inertia. The temperatures of the plates are controlled to $\pm 0.3\,\degr$C to minimise temperature-dependent effects. Laboratory measurements indicate that the temperature coefficients of the power meters are $\approx$ 0.025 \% / K.  The temperature coefficients of the N-type splitter and 50 dB attenuator is $<$ 0.01 \% / K.  Since we also measure the temperatures of these components small corrections can be applied when required.

The CW calibration systems inevitably radiate power out of the horns in the protected radio astronomy band (1400 MHz to 1427~MHz) but at a very low level. At the spare polarizer ports, where the CWs are injected, after splitting and attenuation the power levels are -80 dBm. The effect of this emission on the rest of the world is much diminished by the low level of the sidelobes, further mitigated by the large ground screen. L-BASS is situated in the heart of the radio astronomy observatory with active observations taking place in the protected band. Neither the Lovell Telescope or the Mk2 telescope, that is situated $<$ 100~m from L-BASS, are able to detect our CWs.

\subsection{The physical temperature monitoring system }
\label{temp_monitor}
\begin{figure*}   
\centering
\includegraphics[scale=0.8]{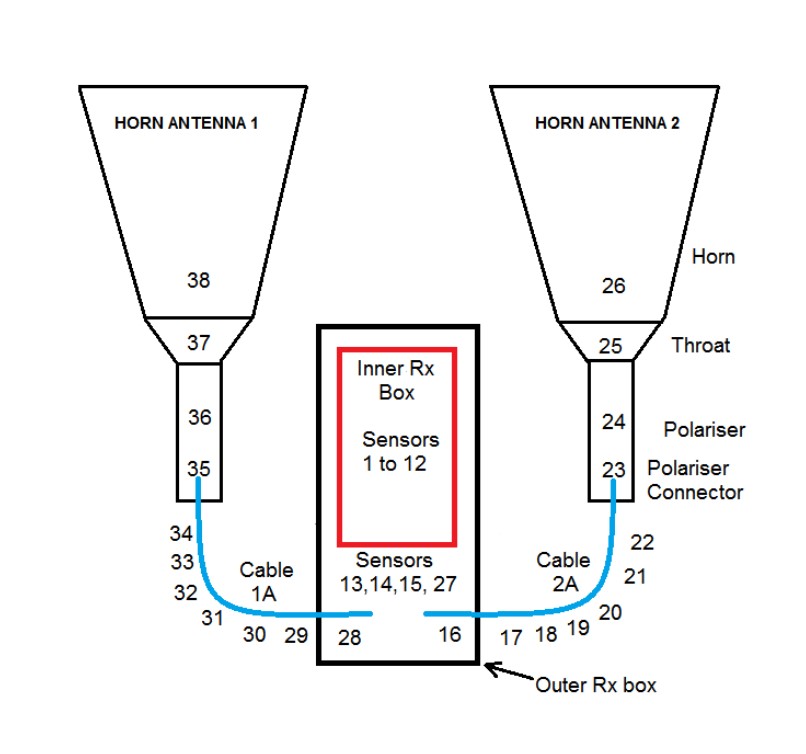}
\caption{Schematic (not to scale) overview of the positions of the temperature sensors on the L-BASS instrument. Sensors 1-12 monitor active components in the receiver; sensors 13-15 and 27 are placed on the ports of the first and second hybrids; sensors 28-38 and 16-26 are fixed  to the passive components ahead of the active receiver – see also Fig.~\ref{horn-and-polariser}.} 
\label{sensors} 
\end{figure*}
Physical temperature measurement and monitoring of the whole L-BASS system is vitally important if we are to achieve our principal science goal of an absolute radiometric accuracy of 0.1~K. As we have already noted, the characteristics of all individual components are temperature dependent; firstly in particular the gains and noise temperature of the LNAs and secondly the losses in the passive components before the LNAs viz the horns, polarizers and RF cables. The largest changes can be expected in the 4.04~m RF cables, connecting the horns to the first hybrid in the receiver box which have ohmic losses of $\approx 0.74$~dB.  With sufficient temperature measurements along the cables the variations in their loss and hence their thermal emission can be accurately modelled and corrections can then be made to the overall system noise budget. This `passives model' also takes into account significant contributions from the losses in the polarizers on each horn. More details of this modelling are given in Paper III.

The temperature monitoring system used is based on a 1-Wire product manufactured by Maxim Integrated Inc.   1-Wire networks are an arrangement of devices, communication lines and connections, with user defined topologies and hardware components.  The L-BASS system comprised digital temperature sensors, a USB-to-1-Wire adapter and appropriate connectors  (all from Maxim Integrated Inc) linked together with 30 AWG gauge wire. Data from the sensors are gathered via the USB-to-1-Wire Adapter and a Raspberry Pi\,3 (model B+) computer running our bespoke Python software; the Raspberry Pi applies time stamps  synchronised to UTC every minute; the data  are output in CSV format and read into the master PC used for data logging.

The temperature sensors are Maxim Integrated model DS18B20, each has a unique 64-bit serial code, thus allowing multiple units to function through the same 1-Wire bus. The sensors have a default resolution of 12-bits corresponding to temperature quantisation of 0.0625\,\degr C.  Our tests showed that at 0\,\degr C they have essentially zero offset.  A linear network topology was used in which the Data, Ground and supply power wires start from the 1-Wire USB adapter bus master and extend to the furthest slave device; the total length of the network is 23\,m.  The schematic diagram in Figure~\ref{sensors} is an overview of the general positions of the 38 1-Wire DS18B20 thermometers attached to the L-BASS instrument. 
Full  details of the temperature monitoring system and its extensive testing prior to deployment on L-BASS are given by \citet{Zerafa_2022}. 

\subsection{The data logging system}
\label{sec:data_logging}
Of the raw 450\,MHz output of the spectrometer,
39.4\,MHz (718 channels) are archived to FITS binary data files at the raw 0.1\,s time sampling rate. The data are timestamped in UTC. The phase switch pulse state is also recorded for each sample. Each file contains one hour of data to keep data loss to $< 1$\,hr in the very rare event of a system crash. The FITS files also archive ancillary data including the 1-wire temperatures, the CW total power monitoring, and meteorological data, in a separate extension table sampled once per minute. All data are also archived on a Network-Attached Device (NAS drive).

The FITS writing program also produces real-time plots of the integrated in-band power of the four output channels (2 IF channels $\times$
2 phase-switch states) vs time.  An example of the spectra of these four outputs is illustrated in Figure 4 of Paper I.\\

The raw FITS files are routinely processed with custom data handling and analysis software \citep{Black_2024} to:
\begin{enumerate}
    \item separate the two phase-switch states from each output of the receiver;
    \item apply a preliminary bandpass calibration to flatten the band;
    \item apply  preliminary gain normalisation to the four outputs to allow for small differences in gain;
    \item evaluate the CW amplitude by subtracting the continuum baseline from the frequency bin(s) containing the CW signal;
    \item interpolate over the frequency bin(s) containing CW signals so that the CW signal does not alter our 1400--1425\,MHz measure of the continuum power;
    \item separate the \ion{H}{i} spectrum over
    $|v_{\rm LSRK}| \pm 200 {\rm \, km\, s}^{-1}$, and replace it with an interpolation over the gap;
    \item average the total powers over the band 1400--1425\,MHz.
\end{enumerate}
The next stages of processing, including differencing and the application of the CW calibration, will be described in Paper III.

\section{Summary}
We have described the components and sub-systems of the dual antenna L-BASS system which will be used to make precise radiometric temperature measurements of the sky relative to the temperature of the North Celestial Pole. The properties of the critical components have been characterized with laboratory measurements, particularly their temperature dependence, so that where necessary temperature stabilization could be implemented and temperatures monitored enabling retrospective corrections to be applied. The system is assembled and final commissioning work is underway. In Paper III (Black et al. in preparation) we will provide a quantitative description of the calibration of the entire system.  A further paper describing the cryogenically cooled load which will be used to establish the absolute temperature scale will follow in due course. 

\section*{Acknowledgements}

DZ acknowledges support from an STFC studentship. Construction of L-BASS was part-funded by STFC grant ST/L000768/1. We thank Ralph Spencer for designing the phase switch driver circuit. Without the sterling efforts of the Jodrell Bank technical staff, particularly Jon Edgley, Adrian Galtress, Dave Clark and Alan Williams, the project would not have been possible. Eddie Blackhurst is thanked for his role in testing the receiver and tackling cross-talk issues.  We also thank Jordan Norris for his work to implement the CW calibration system and Joe Metson and Louis Hurrey for their work on the solar observations. PJB acknowledges support from the UoM Department of Physics \& Astronomy via a Department Scholarship.
Finally we would like to thank the anonymous reviewers for an encouraging and constructive appraisal of this work. Feedback about receiver chain component properties/functions, the uncertainties inherent in measured quantities and clarity of technical diagrams have enhanced the presentation of this Paper. 

\section*{Data Availability Statement}

The instrument design and testing data underlying this article are either available in \cite{Zerafa_2022} in the first instance or otherwise by written request to the corresponding author.

 



\bibliographystyle{mnras}
\bibliography{lbassbiblio} 

\begin{thebibliography}{}
\makeatletter
\relax
\def\mn@urlcharsother{\let\do\@makeother \do\$\do\&\do\#\do\^\do\_\do\%\do\~}
\def\mn@doi{\begingroup\mn@urlcharsother \@ifnextchar [ {\mn@doi@} {\mn@doi@[]}}
\def\mn@doi@[#1]#2{\def\@tempa{#1}\ifx\@tempa\@empty \href {http://dx.doi.org/#2} {doi:#2}\else \href {http://dx.doi.org/#2} {#1}\fi \endgroup}
\def\mn@eprint#1#2{\mn@eprint@#1:#2::\@nil}
\def\mn@eprint@arXiv#1{\href {http://arxiv.org/abs/#1} {{\tt arXiv:#1}}}
\def\mn@eprint@dblp#1{\href {http://dblp.uni-trier.de/rec/bibtex/#1.xml} {dblp:#1}}
\def\mn@eprint@#1:#2:#3:#4\@nil{\def\@tempa {#1}\def\@tempb {#2}\def\@tempc {#3}\ifx \@tempc \@empty \let \@tempc \@tempb \let \@tempb \@tempa \fi \ifx \@tempb \@empty \def\@tempb {arXiv}\fi \@ifundefined {mn@eprint@\@tempb}{\@tempb:\@tempc}{\expandafter \expandafter \csname mn@eprint@\@tempb\endcsname \expandafter{\@tempc}}}

\bibitem[\protect\citeauthoryear{Behe \& Brachat}{Behe \& Brachat}{1991}]{Behe1991}
Behe R.,  Brachat P.,  1991, \mn@doi [IEEE Transactions on Antennas and Propagation] {10.1109/8.97358}, 39, 1222

\bibitem[\protect\citeauthoryear{{Bennett} et~al.,}{{Bennett} et~al.}{2013}]{Bennett2013}
{Bennett} C.~L.,  et~al., 2013, \mn@doi [\apjs] {10.1088/0067-0049/208/2/20}, \href {https://ui.adsabs.harvard.edu/abs/2013ApJS..208...20B} {208, 20}

\bibitem[\protect\citeauthoryear{{Bersanelli} et~al.,}{{Bersanelli} et~al.}{2010}]{2010A&A...520A...4B}
{Bersanelli} M.,  et~al., 2010, \mn@doi [\aap] {10.1051/0004-6361/200912853}, \href {https://ui.adsabs.harvard.edu/abs/2010A&A...520A...4B} {520, A4}

\bibitem[\protect\citeauthoryear{Black}{Black}{2024}]{Black_2024}
Black P.~J.,  2024, Master's thesis, University of Manchester, Jodrell Bank Centre for Astrophysics, United Kingdom, \url {https://research.manchester.ac.uk/en/studentTheses/commissioning-and-data-handling-for-the-l-bass-instrument}

\bibitem[\protect\citeauthoryear{{Calabretta}, {Staveley-Smith}  \& {Barnes}}{{Calabretta} et~al.}{2014}]{Calabretta2014}
{Calabretta} M.~R.,  {Staveley-Smith} L.,   {Barnes} D.~G.,  2014, \mn@doi [\pasa] {10.1017/pasa.2013.36}, \href {https://ui.adsabs.harvard.edu/abs/2014PASA...31....7C} {31, e007}

\bibitem[\protect\citeauthoryear{{Dassault Systemes}}{{Dassault Systemes}}{2017}]{CST_STUDIO}
{Dassault Systemes} 2017, CST Studio Suite, \url {https://www.3ds.com/products/simulia/cst-studio-suite}

\bibitem[\protect\citeauthoryear{{Fixsen} et~al.,}{{Fixsen} et~al.}{2011}]{Fixsen2011}
{Fixsen} D.~J.,  et~al., 2011, \mn@doi [\apj] {10.1088/0004-637X/734/1/5}, \href {http://adsabs.harvard.edu/abs/2011ApJ...734....5F} {734, 5}

\bibitem[\protect\citeauthoryear{Jarosik et~al.,}{Jarosik et~al.}{2003}]{Jarosik_2003}
Jarosik N.,  et~al., 2003, \mn@doi [\apjs] {10.1086/346080}, 145, 413

\bibitem[\protect\citeauthoryear{{Leech} et~al.,}{{Leech} et~al.}{2011}]{Leech2011}
{Leech} J.,  et~al., 2011, \mn@doi [\aap] {10.1051/0004-6361/201117124}, \href {https://ui.adsabs.harvard.edu/abs/2011A&A...532A..61L} {532, A61}

\bibitem[\protect\citeauthoryear{{Mician GmbH}}{{Mician GmbH}}{2015}]{Microwave_Wizard}
{Mician GmbH} 2015, Microwave Wizard V.8, \url {https://www.mician.com}

\bibitem[\protect\citeauthoryear{Pollak, Holler, Jones  \& Taylor}{Pollak et~al.}{2019}]{Pollak_2019}
Pollak A.~W.,  Holler C.~M.,  Jones M.~E.,   Taylor A.~C.,  2019, \mn@doi [\mnras] {10.1093/mnras/stz2169}, 489, 548

\bibitem[\protect\citeauthoryear{{Radiometer Physics GmbH}}{{Radiometer Physics GmbH}}{2013}]{labview}
{Radiometer Physics GmbH} 2013, LABVIEW, \url {https://www.radiometer-physics.de}

\bibitem[\protect\citeauthoryear{{Reich}, {Reich}  \& {Testori}}{{Reich} et~al.}{2004}]{Reich2004}
{Reich} P.,  {Reich} W.,   {Testori} J.~C.,  2004, in {Uyaniker} B.,  {Reich} W.,   {Wielebinski} R.,  eds, The Magnetized Interstellar Medium. pp 63--68

\bibitem[\protect\citeauthoryear{{Seiffert}, {Mennella}, {Burigana}, {Mandolesi}, {Bersanelli}, {Meinhold}  \& {Lubin}}{{Seiffert} et~al.}{2002}]{Seiffert2002}
{Seiffert} M.,  {Mennella} A.,  {Burigana} C.,  {Mandolesi} N.,  {Bersanelli} M.,  {Meinhold} P.,   {Lubin} P.,  2002, \mn@doi [\aap] {10.1051/0004-6361:20020880}, \href {https://ui.adsabs.harvard.edu/abs/2002A&A...391.1185S} {391, 1185}

\bibitem[\protect\citeauthoryear{{Seiffert} et~al.,}{{Seiffert} et~al.}{2011}]{Seiffert2011}
{Seiffert} M.,  et~al., 2011, \mn@doi [\apj] {10.1088/0004-637X/734/1/6}, \href {http://adsabs.harvard.edu/abs/2011ApJ...734....6S} {734, 6}

\bibitem[\protect\citeauthoryear{{Wolleben} et~al.,}{{Wolleben} et~al.}{2021}]{Wolleben2021}
{Wolleben} M.,  et~al., 2021, \mn@doi [\aj] {10.3847/1538-3881/abf7c1}, \href {https://ui.adsabs.harvard.edu/abs/2021AJ....162...35W} {162, 35}

\bibitem[\protect\citeauthoryear{Zerafa}{Zerafa}{2022}]{Zerafa_2022}
Zerafa D.,  2022, PhD thesis, University of Manchester, Jodrell Bank Centre for Astrophysics, United Kingdom, \url {https://research.manchester.ac.uk/en/studentTheses/evaluating-the-temperature-of-the-sky-the-l-band-all-sky-survey}

\bibitem[\protect\citeauthoryear{Zerafa, Wilkinson, Radcliffe, Leahy, Browne  \& Black}{Zerafa et~al.}{2025}]{Paper1}
Zerafa D.,  Wilkinson P.~N.,  Radcliffe C.~J.,  Leahy J.~P.,  Browne I. W.~A.,   Black P.~J.,  2025, \mn@doi [RASTI] {https://doi.org/10.1093/rasti/rzaf017}, accepted

\makeatother
\end{thebibliography}





\bsp	
\label{lastpage}
\end{document}